\documentclass[journal]{IEEEtran}
\IEEEoverridecommandlockouts
\usepackage[utf8]{inputenc}
\usepackage{cite,balance}
\usepackage{amsmath,amssymb,amsfonts,multirow,upgreek,subcaption}
\usepackage[table]{xcolor}
\usepackage{textcomp}
\usepackage{xcolor}
\usepackage{bm}
\usepackage{algpseudocode}
\usepackage{algorithm}
\usepackage{enumitem}
\usepackage{psfrag}
\usepackage{tikz}
\usetikzlibrary{shapes.geometric, arrows}
\tikzstyle{roundrec} = [font=\footnotesize,rectangle, rounded corners, minimum width=1cm, minimum height=1cm,text centered, text width=2.5cm, draw=black]
\tikzstyle{rec} = [font=\footnotesize,rectangle, minimum width=3cm, minimum height=1cm, text width=3.75cm, draw=black]
\tikzstyle{bubble} = [font=\footnotesize,rectangle, rounded corners, text centered, draw=black]
\tikzstyle{arrow} = [thick,->,>=stealth]
\def\BibTeX{{\rm B\kern-.05em{\sc i\kern-.025em b}\kern-.08em
    T\kern-.1667em\lower.7ex\hbox{E}\kern-.125emX}}

\renewcommand{\j}{\mathrm{j}}
\DeclareMathOperator*{\argmax}{arg\,max}

\newcommand*{\herm}{^{\mathsf{H}}}
\newcommand*{\transp}{^{\mathsf{T}}}

\title{Adaptive Downlink Localization and User Tracking in Near-Field and Far-Field: A Trade-Off Analysis}
\author{Georgios Mylonopoulos, Behrooz Makki, \emph{Senior Member IEEE}, Stefano Buzzi, \emph{Senior Member IEEE}, G\'abor Fodor, \emph{Senior Member IEEE} \vspace{-1cm}
\thanks{This work has received funding from the European Union’s Horizon 2020 research and innovation
programme under the Marie Skłodowska-Curie grant agreement No 956256. \\
Georgios Mylonopoulos is with the Department of Electrical and Information Engineering, University of Cassino and Southern Lazio, 03043 Cassino, Italy,
and also with the Consorzio Nazionale Interuniversitario per le Telecomunicazioni (CNIT), 43124 Parma, Italy and also with Ericsson Research, Ericsson, 417 56 Göteborg, Sweden (e-mail: georgios.mylonopoulos@unicas.it). \\
Behrooz Makki is with Ericsson Research, Ericsson, 417 56 Göteborg,
Sweden (e-mail: behrooz.makki@ericsson.com). \\
Stefano Buzzi is with the Department of Electrical and Information Engineering, University of Cassino and Southern Lazio, 03043 Cassino, Italy,
and also with the Consorzio Nazionale Interuniversitario per le Telecomunicazioni (CNIT), 43124 Parma, Italy. He is also affiliated with Politecnico di Milano, Milan, Italy (e-mail: buzzi@unicas.it). \\
Gábor Fodor is with Ericsson Research, 16480 Stockholm, Sweden, and
also with the Division of Decision and Control, KTH Royal Institute of
Technology, 11428 Stockholm, Sweden (e-mail: gabor.fodor@ericsson.com). \\
A short version of this paper has been submitted for possible presentation at ICC Workshops’ 2024. A pre-print is available on ArXiv~\cite{ArxivAdaptiveDL}.}
}
\date{January 2023}

\begin{document}

\maketitle

\begin{abstract}
This paper considers the problem of downlink localization and user equipments (UEs) tracking with an adaptive procedure. Our proposed scheme addresses the accuracy-complexity trade-off in detection, localization and tracking of UEs in a broad range of distances. We provide the base station (BS) with two different signaling schemes and the UEs with two localization algorithms; one assuming far-field (FF) and the other assuming near-field (NF) conditions. Using the proposed signaling schemes allows for different beam-sweep patterns, where their geometrical compatibility depends on the BS-to-UE distance. In that sense, the FF-NF distinction transcends the traditional definition of FF and NF propagation, which solely refers to the spherical nature of the impinging wavefronts. Specifically, our proposed NF scheme requires beams focused on specific spots and thus more transmissions are required to sweep the area, while the FF scheme assumes distant UEs and fewer beams are sufficient to cover different directions. Moreover, the UEs need to employ a localization algorithm adapted for the employed signaling scheme at the BS side. Since in the proposed scheme the validity of the FF and NF assumptions must be communicated by the UEs back to the BS, we derive a low-complexity algorithm that exploits the simplified FF channel model and highlight the practical benefits and the limitations of this assumption. Also, we propose multiple iterations of the localization procedure and an adaptive model, where an appropriate signaling scheme is used depending on the expected accuracy-complexity trade-off. Multiple localization iterations introduce a sense of memory and allow us to consider a tracking application, where the formed trajectory serves as a metric for the validity of the FF and NF assumptions. Moreover, for the inspected geometry, the range from the BS, where the FF signaling scheme can be successfully employed, is investigated. We show that relying on the conventional Fraunhofer distance is not sufficient for adapting localization and tracking algorithms in the mixed NF and FF environment, as it fails to encapsulate the signaling requirements of each localization scheme.
\end{abstract}

\section{Introduction}

Localization and spatial awareness will play an important role in 6G networks with high accuracy positioning and sensing providing several opportunities and challenges for enhanced network performance~\cite{de2021convergent}. The ability of a localization procedure to extract spatial information is strongly tied to the employed channel model and its ability to accurately describe the propagation environment. The near-field (NF) and far-field (FF) channel model mismatch is analyzed in~\cite{chen2022channel} for extremely large antenna arrays, from a spatial information perspective. In 5G networks the FF assumption is consistent, but larger arrays, higher frequency bands and novel hardware developments in 6G networks, such as reconfigurable intelligent surfaces (RISs), will lead to more practical scenarios where the users (UEs) may be in the base station’s (BS) NF region~\cite{cui2022near}. 

The problem of localization in the NF has gained a lot of interest with several techniques employed to account for the complex NF channel model. NF source localization is explored in~\cite{jingjing2021search}, with an error correction procedure that employs the exact signal model after an initial relaxed parameter estimation. In~\cite{su2021deep}, a deep unfolding network is employed for the angular and range decoupling for the NF source localization problem. The additional degrees of freedom of utilizing nested arrays in localization procedures is explored in~\cite{shu2020near} for NF sources, despite the non-uniform behavior introduced by the spherical wavefront. The fourth order cumulant may decouple the angular domain information for NF signal classification and~\cite{guanghui2019high} presents a high accuracy localization algorithm. In addition, localization utilizing a receive strength indicator (RSSI) in the angular domain is explored in~\cite{huang2022near} with an RIS, while the concept of FF and NF classification is considered. Similarly, multiple RISs are considered in~\cite{ArxvivUEDetRIS} to jointly detect active UEs and localize them in the NF. 

The process of localization is often one component of a multi-variable problem. The joint problem of communication and localization is explored in~\cite{yang2021communication}, where the UEs are in the NF region. Also, the joint localization and channel estimation in an RIS-assisted system is explored in~\cite{pan2023ris}, under the NF assumption due to the extra-large size of the RIS. The target localization within the NF region via indirect paths, provided by an RIS and the RIS phase design that accounts for the spherical nature of the impinging wavefront are inspected in~\cite{luan2021phase}. 

Compared to the true model, the FF assumption provides a simpler channel model that allows for simple positioning algorithms. The concept of downlink (DL) localization is explored in~\cite{fascista2021downlink}, assuming FF propagation, with a formal maximum likelihood (ML) solution and a Cramér-Rao analysis. The single BS-single snapshot DL localization is extended to the 3-dimensional, asynchronous MIMO case in~\cite{nazari2023mmwave} with the corresponding Fisher information analysis. Exploring the virtual anchor provided by an active RIS, a practical localization scheme is introduced in~\cite{mylonopoulos2022active} with a sweeping beam-steering design for FF propagation, while the ML and other practical relaxed estimators are derived in~\cite{mylonopoulos2023maximum}.

As the border of the NF region increases, a mixed model system needs to be investigated. The UE communication problem in the NF-FF border is explored in~\cite{de2020near}, with the exact NF channel model being essential as the UE approaches the large array. Similar to communication, positioning accuracy may benefit from NF propagation. There are several beam-forming challenges that need to be addressed as NF beam-focusing and traditional FF beam-steering techniques are inherently different~\cite{zhang20236g}. The effectiveness of FF beamforming in the NF region and the limitations of the array gain relative to the Fraunhofer distance (FD) are explored in~\cite{bjornson2021primer}. In~\cite{he2021mixed,tian2021phase,huang2020one,wang2017unified,he2022mixed,yan2023improved}, radio source localization has been explored in a mixed FF and NF model with spherical and planar waves simultaneously received. Separating the FF and NF sources is investigated in~\cite{tian2021phase}, while a localization algorithm, based on peak-extraction for the estimated parameters, is presented in~\cite{huang2020one}. In~\cite{wang2017unified} the localization is unified in the angular domain of a source that is either in the FF or the NF region.  

Localization procedures benefit from multiple iterations as more spatial information is extracted and the joint process of prior estimations introduces a sense of memory that allows for tracking applications to emerge. In~\cite{guerra2021near}, the phase difference profile for large arrays is analyzed, deriving practical algorithms and fundamental limits in an, RIS-aided, UE tracking application that considers NF propagation conditions. The precoding design and the RIS optimization of a similar tracking application is investigated in~\cite{palmucci2023two}. Finally, an adaptive localization procedure is introduced in~\cite{li2020self}, where the adaptive component is designed to handle unreliable estimations that decrease the system’s robustness.

In this paper, we develop an adaptive DL localization and tracking procedure composed of two distinctive signaling schemes and localization algorithms, based on the validity of the FF and the NF assumptions. Our FF-based scheme assumes a large BS-to-UE distance to exploit the relaxed channel model and a beam-steering technique sweeps the reduced angular domain, resulting in lower signaling overhead and computational complexity. On the other hand, our NF-based scheme always considers the true channel model and extensively covers the whole region utilizing beam-focusing technique. With this intuition, our proposed adaptive signaling scheme poses an accuracy-complexity trade-off, where the UEs are able to communicate to the BS whether the low-complexity, FF-based, scheme can accurately be employed or not. Moreover, based on the developed FF- and NF-based scheme, we develop an iterative algorithm for UE tracking. Finally, we study the range in which the NF- and FF-based schemes can be successfully employed, and compare it with the traditional FD boundary. Our proposed adaptive scheme address the accuracy-complexity trade-off for detection, localization and tracking of UEs in a broad range of distances. The considered scenario is distinctive, as the assumed channel model is not unique, but rather obtained from the BS-to-UE distance. We perform the complexity analysis of our proposed schemes and evaluate their performance for different parameter settings.

Our results show that our proposed adaptive localization scheme balances the positioning accuracy and tracking abilities of the system with acceptable signaling overhead and computational requirements. Moreover, the switching distance between the FF and the NF schemes diverges from the FD and depends, among other factors, on the considered geometry and beam-forming techniques employed with each signaling scheme, as well as the computational complexity. Jointly considering past and current positional information offers higher accuracy and a reliable indicator for the adaptive signaling requirements.

The main contributions of our paper may be summarised as follows: 
\begin{itemize}
\item We study the performance limitations of a DL positioning algorithm, designed around the FF channel model, while the UE is in the NF.
\item We highlight the algorithmic benefits that the FF channel model provides in terms of positioning, compared to the NF model.
\item We explore the low signaling requirements in a DL localization scheme, when the UE is not close to the BS, in terms of beamforming and spatial energy distribution. 
\item We explore the increased signaling requirements in a DL localization scheme, when the UE is close to the BS, in terms of beamforming and spatial energy distribution. 
\item We impose an accuracy-to-resources trade-off between the FF-based and the NF-based localization schemes. The characteristics of the imposed trade-off change, as the BS-to-UE distance changes.
\item We emphasize the advantages of an adaptive cooperation of FF and NF in a DL localization scheme, in terms of signaling overhead, computational complexity and performance, as we highlight the flexibility in combining the two schemes.
\item We investigate the FD as a practical junction point for the proposed cooperation of the FF and NF schemes and we illustrate its limitations, as the signaling design and the considered geometry better encapsulate the proper combination of the two schemes.
\item We introduce a sense of memory in our system as we introduce a tracking application. Thus, we illustrate that prior information and spatial awareness provide a strong and versatile adaptive protocol that effectively balances the accuracy-to-resources trade-off, imposed by our two proposed schemes. 
\end{itemize}

\emph{Notation}: Column vectors and matrices are denoted by lowercase and uppercase boldface letters, respectively. The symbols $(\cdot)\transp$ and $(\cdot)\herm$ denote transpose and conjugate-transpose and Kronecker product, respectively. Also, $\left \| \bm{\alpha} \right \|$ is the Euclidean norm of the vector $ \bm{\alpha} $ and $\rm{card}\{\mathcal{M}\}$ counts the number of elements within the $\mathcal{M}$ set, while $\j$ is the imaginary unit.
\begin{figure*}[h]
    \centering
    \includegraphics[width=\linewidth]{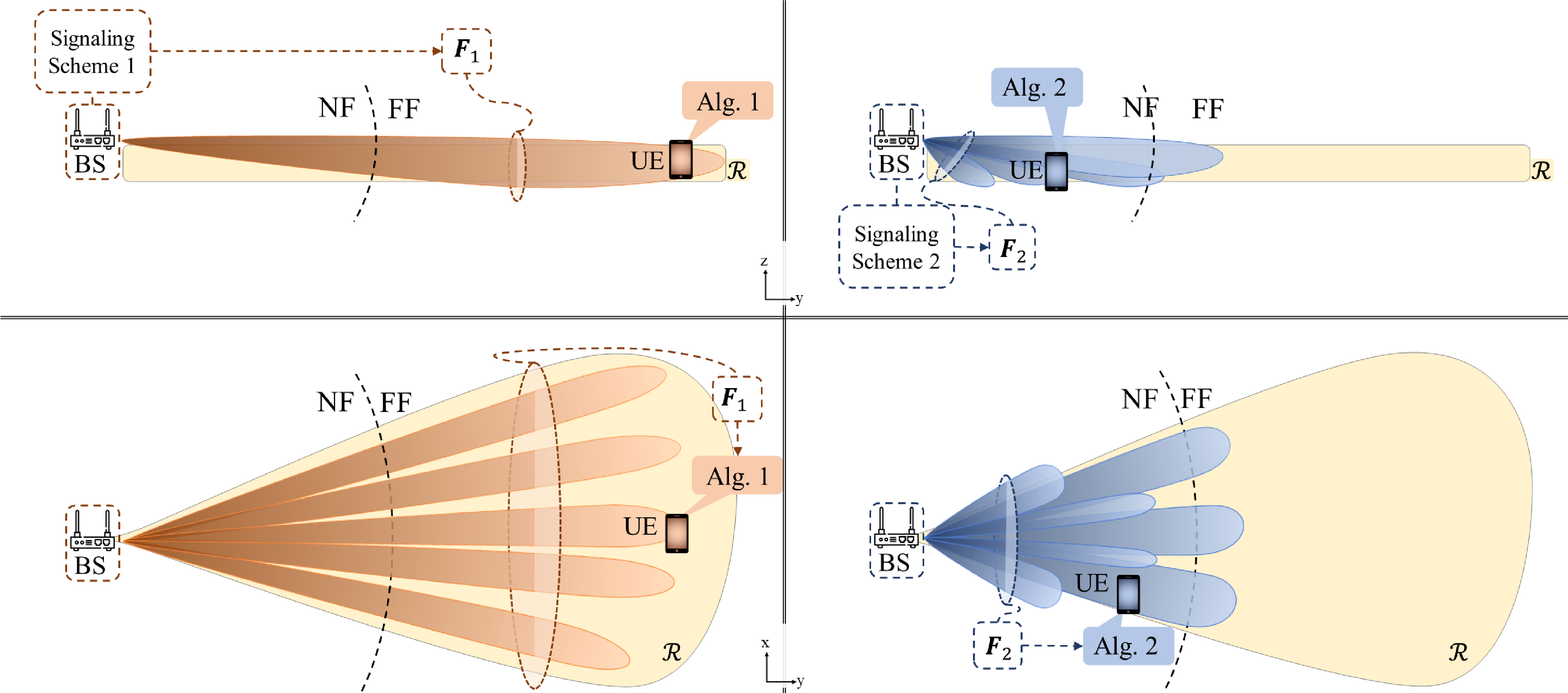}
    \caption{Considered scenario where BS provides 2 signaling options, designed for DL UE localization in the FF and the NF, respectively. The UE may employ a localization algorithm fit for the corresponding localization scheme.}
    \label{fig:sm}
\end{figure*}

\section{System Model}
In this section, we introduce the considered system-scenario and explore the dual channel model and  dual beam design strategies that correspond to the FF or NF assumptions. Figure~\ref{fig:sm} illustrates the geometrical setup. There is a wall mounted BS at point $\bm{p}_{\rm{BS}}$ equipped with a uniform planar array (UPA) of $N_{\rm{BS}} = N_{\rm{BS}}^{(\rm{x})} \times N_{\rm{BS}}^{(\rm{z})}$ antennas with an inter-antenna spacing of $\delta_{\rm{BS}}$ in both dimensions, serving single antenna UEs within the region $\mathcal{R}$. The UEs need to estimate their own position, $\bm{p}_{\rm{UE}}$ and communicate it to the BS, where the localization procedure is orchestrated. 

\begin{figure}
    \centering
    \begin{tikzpicture}[node distance=2cm]
        \node (start) at (1.5,9.5) [roundrec] {The BS employs one \emph{Signaling Scheme}};
        \node (scheme1) at (-0.9,8.25) [rec] {\textbf{FF-based signaling}:\\\textbullet The BS uses $\bm{F}_{1}$.\\\textbullet The UEs use Alg. 1 for $\hat{\bm{p}}_{\rm{UE}}$.};
        \node (scheme2) at (3.85,8.25) [rec] {\textbf{NF-based signaling}:\\\textbullet The BS uses $\bm{F}_{2}$.\\\textbullet The UEs use Alg. 2 for $\hat{\bm{p}}_{\rm{UE}}$.};
        \node (dec) at (-0.9,7.1) [bubble] {$\hat{\bm{p}}_{\rm{UE}}$ in FF};
        \node (end) at (1.5,5.75) [bubble] {Process completed};
        \node (init) at (1.5,10.5) [bubble] {Initiate Process};
        
        \draw [arrow] (node cs:name=init,anchor=south) -- (node cs:name=start,anchor=north);
        \draw [arrow] (node cs:name=start,anchor=west) --(-0.9,9.5)-- node[xshift=-0.5cm,yshift=0.6cm] {\footnotesize BS opts for scheme 1}(node cs:name=scheme1,anchor=north);
        \draw [arrow] (node cs:name=start,anchor=east) --(3.85,9.5)-- node[xshift=0.5cm,yshift=0.6cm] {\footnotesize BS opts for scheme 2}(node cs:name=scheme2,anchor=north);
        \draw [arrow] (node cs:name=scheme1,anchor=south) -- (node cs:name=dec,anchor=north);
        \draw [arrow] (node cs:name=dec,anchor=west) --node[anchor=north] {\footnotesize True}(-2.75,7.1)--(-2.75,6.35)--(1.5,6.35)-- (node cs:name=end,anchor=north);
        \draw [arrow] (node cs:name=dec,anchor=east) --node[anchor=north] {\footnotesize False}(1.5,7.1)--(1.5,8.25)-- (node cs:name=scheme2,anchor=west);
        \draw [arrow] (node cs:name=scheme2,anchor=south) --(3.85,6.35)--(1.5,6.35)-- (node cs:name=end,anchor=north);
    \end{tikzpicture}
    \caption{Proposed DL localization signaling procedure.}
    \label{fig:loc_proc}
\end{figure}
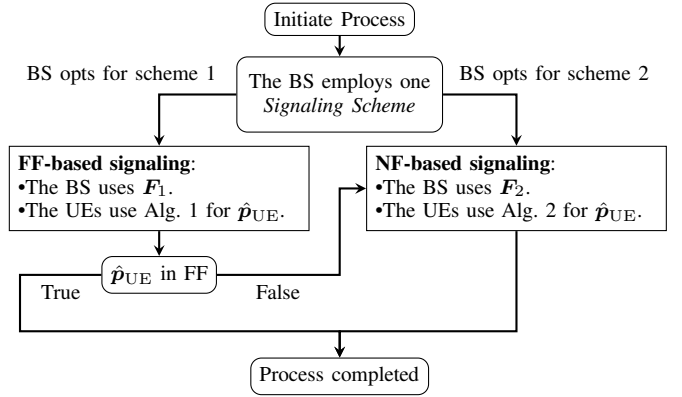

The localization procedure is summarised in Fig.~\ref{fig:loc_proc}, with two distinctive signaling schemes available to the BS:
\begin{itemize}
\item \emph{FF-based signaling}: Under the FF assumption, the BS transmits $J_{1}$ OFDM pilots, employing the beam-steering matrix $\bm{F}_{1} \in \mathbb{C}^{N_{\rm{BS}}\times J_{1}}$. The design of $\bm{F}_{1}$ exploits geometrical generalizations for $\mathcal{R}$, such that the transmitted pilots are reduced.
\item \emph{NF-based signaling}: Under the NF assumption, the BS transmits $J_{2}$ OFDM pilots, employing the beam-focusing matrix $\bm{F}_{2} \in \mathbb{C}^{N_{\rm{BS}}\times J_{2}}$. The design of $\bm{F}_{2}$ is more complex with more spatial beams, such that $\mathcal{R}$ is adequately covered.
\end{itemize}
The OFDM pilots consist of $Q$ subcarriers with a subcarrier bandwidth (BW) of $W_{\rm{o}}$ and a subcarrier spacing of $W_{\rm{sub}}$. The total BW is $(Q-1)W_{\rm{sub}} + W_{\rm{o}}$.

\subsection{Channel \& Signal Model}
Here, we present the true geometrical channel model and explore a relaxed model that assumes a large BS-to-UE distance. The BS is assumed to have a strong line-of-sight (LOS) link with all UEs. The true channel between the BS and the $m^{\rm{th}}$ UE for the $q^{\rm{th}}$ subcarrier is given by
\begin{subequations}
\begin{align}
\bm{H}_{\rm{NF}}^{(m)} &= [\bm{h}_{1}^{(m)} \dotsc \bm{h}_{q}^{(m)} \dotsc \bm{h}_{Q}^{(m)}] \in \mathbb{C}^{N_{\rm{BS}}\times Q} , \\
\bm{h}_{q}^{(m)} &= \beta^{(m)} \bm{\alpha}_{q}(\bm{p}_{\rm{UE}}^{(m)}) \in \mathbb{C}^{N_{\rm{BS}}\times 1} ,
\end{align}
\label{eq:channel_h_NF}
\end{subequations}
where $\bm{\alpha}_{q}(\bm{p})$ is the steering vector for point $\bm{p}$, relative to the BS and $\beta^{(m)}$ is a scalar accounting for the free-space path loss and the synchronization error for the $m^{\rm{th}}$ UE, given by
\begin{align}
\beta^{(m)} =  \frac{\lambda_{\rm{o}}}{4\pi d^{(m)}} \text{exp}\bigl\{-\j \phi_{\rm{o}}^{(m)}\bigr\}.
\label{eq:beta_qm}
\end{align}
Here, $\lambda_{\rm{o}}$ is the wavelength for the considered frequency, $d^{(m)} = \| \bm{p}_{\rm{BS}} - \bm{p}_{\rm{UE}}^{(m)}\|$ is the distance of flight (DOF) for the $m^{\rm{th}}$ UE relative to the BS's point of reference, $\bm{p}_{\rm{BS}}$, and $\phi_{\rm{o}}$ a uniformly random phase offset, accounting for the synchronization mismatch. The $n^{\rm{th}}$ element of the steering vector, $\bm{\alpha}_{q}(\bm{p})$, is  
\begin{align}
\bigl[ \bm{\alpha}_{q}(\bm{p})\bigr]_{n} = \frac{\lambda_{q}d^{(m)}}{\lambda_{\rm{o}}\|\bm{p} - \bm{p}_{\rm{BS}}^{(n)} \|}\text{exp}\bigl\{ -\j\frac{2\pi}{\lambda_{q}} \|\bm{p} - \bm{p}_{\rm{BS}}^{(n)} \|\bigr\} \quad , 
\label{eq:sv_NF}
\end{align}
where $\bm{p}_{\rm{BS}}^{(n)}$ is the position of the $n^{\rm{th}}$ BS antenna, for $n = 1,\dots,N_{\rm{BS}}$. The channel in~\eqref{eq:channel_h_NF} preserves the spherical wavefront as it makes no assumptions and it is essentially the NF channel model. On the other hand, assuming a large BS-to-UE distance, an approximation of~\eqref{eq:channel_h_NF} is given by 
\begin{subequations}
\begin{align}
\tilde{\bm{h}}_{q}^{(m)} &= \beta^{(m)}\text{exp}\bigl\{-\j2\pi(q-1)\frac{W_{\rm{sub}}}{c_{\rm{o}}}d^{(m)}\bigr\}\tilde{\bm{\alpha}}(\bm{\theta}_{\rm{UE}}^{(m)}), \\
\bm{H}_{\rm{FF}}^{(m)} &= [\tilde{\bm{h}}_{1}^{(m)} \dotsc \tilde{\bm{h}}_{q}^{(m)} \dotsc \tilde{\bm{h}}_{Q}^{(m)}] \quad , \notag \\
&= \beta^{(m)}\bm{t}\transp(d^{(m)})\tilde{\bm{\alpha}}(\bm{\theta}_{\rm{UE}}^{(m)}) \in \mathbb{C}^{N_{\rm{BS}}\times Q} ,
\end{align}
\label{eq:channel_h_FF}
\end{subequations}
where $\tilde{\bm{\alpha}}(\bm{\theta})$ is the planar wave steering vector for an angle-of-departure (AOD), $\bm{\theta} = [\theta^{\rm{az}}, \theta^{\rm{el}}]$ and 
\begin{align}
\bigl[ \bm{t}(d)\bigr]_{q} = \text{exp}\bigl\{-\j2\pi(q-1)\frac{W_{\rm{sub}}}{c_{\rm{o}}}d\bigr\} \quad .
\label{eq:t}
\end{align}
The relaxed model in~\eqref{eq:channel_h_FF} is essentially the FF model and note that the AOD and the DOF are decoupled.
The received signal for the $j^{\rm{th}}$ spatial stream along the $q^{\rm{th}}$ subcarrier is given by
\begin{align}
y_{q,j}^{(m)} = \sqrt{\mathcal{P}} \bm{f}_{j}\herm\bm{h}_{q}^{(m)}x_{q,j}^{(m)} + z_{q,j} \quad ,
\label{eq:y_qj}
\end{align}
where $\mathcal{P}$ is the transmit power, $\bm{f}_{j}$ is the $j^{\rm{th}}$ precoding vector of the employed signaling scheme, $x_{q,j}^{(m)}$ is the OFDM symbol and $z_{q,j} \sim \mathcal{CN}(0,\sigma_{z}^2)$ denotes the thermal noise. The received signal can be organized in a matrix form, $\bm{Y}\in\mathbb{C}^{J\times Q}$, while the vectors $\bm{y}_{j}\in\mathbb{C}^{J\times 1}$ and $\bm{y}_{q}\in\mathbb{C}^{Q\times 1}$ are the collections of scalars for the $j^{\rm{th}}$ beam and the $q^{\rm{th}}$ subcarrier, respectively.

\subsection{Beam Design}\label{ssec:beam}
Here, we present the core ideas behind the beam design and the process of effectively sweeping $\mathcal{R}$ with the two proposed signaling schemes. Considering Fig.~\ref{fig:sm}, we highlight the signaling overhead advantage of the FF-based approach. For the considered geometry, where 
\begin{align}
\mathcal{R} \quad : \left\{\begin{matrix}
d^{(m)} \in [d_{\rm{min}},d_{\rm{max}}] \\ 
{\theta^{\rm{az}}}^{(m)} \in [\theta^{\rm{az}}_{\rm{min}},\theta^{\rm{az}}_{\rm{max}}] \\ 
 z^{(m)} \in [z_{\rm{min}},z_{\rm{max}}]
\end{matrix}\right.\;, 
\end{align}
has a cylindrical structure with a small span across the z axis and a height misalignment, relative to the BS. The elevation AOD for the $m^{\rm{th}}$ UE is given by 
\begin{align}
{\theta^{\rm{el}}}^{(m)} = \sin^{-1}(\frac{\Delta z^{(m)}}{d^{(m)}}) \; ,
\end{align}
where $\Delta z$ is the height displacement relative to the BS. 
\subsubsection{FF approach}
For the beam-sweep design, the FF assumption refers to a \emph{sufficiently large} DOF. As the DOF increases the angular domain becomes asymptotically one-dimensional, i.e., for a finite $\Delta z$; $\lim_{d\to\infty} \theta_{\rm{el}} = 0$. With this intuition, if $d \gg \Delta z$, we can propose a beam design that sweeps across a single dimension and assume planar waves. Therefore, the number of required spatial beams is reduced and it is dictated by the azimuth span of $\mathcal{R}$. The beam-steering design of the FF-based scheme is simplified to 
\begin{align}
\bm{F}_{1} &= [\tilde{\bm{\alpha}}(\bm{\theta}_{1}) \dotsc \tilde{\bm{\alpha}}(\bm{\theta}_{j}) \dotsc \tilde{\bm{\alpha}}(\bm{\theta}_{J_1})] \; ,\notag \\
\bm{\theta}_{j} &= [\theta_{j}, 0] \;, \quad \theta_{j}\in [\theta^{\rm{az}}_{\rm{min}}, \theta^{\rm{az}}_{\rm{max}}] \quad.
\end{align}
\subsubsection{NF approach}
For the beam-sweep design, the NF assumption refers to a \emph{insufficiently large} DOF. We observe that the FF beam design is problematic as the DOF decreases. The spherical nature of the transmitted wave needs to be addressed and the angular domain is no longer asymptotically one-dimensional, i.e., $\lim_{d\to0} |\theta_{\rm{el}}| = \pi/2$. Therefore, additional spatial beams are required, with a three-dimensional sweep, consisting of a discrete number of focus points. The beam focusing design of the NF-based scheme is  
\begin{align}
\bm{F}_{2} &= [\bm{\alpha}(\bm{p}_{1}) \dotsc \bm{\alpha}(\bm{p}_{j}) \dotsc \bm{\alpha}(\bm{p}_{J_2})] \; , \rm{ where} \notag \\
\bm{p}_{j} &= \bm{p}_{\rm{BS}} + d_{j} \begin{bmatrix}
\cos(\theta^{\rm{el}}_{j})\sin(\theta^{\rm{az}}_{j})\\ 
\cos(\theta^{\rm{el}}_{j})\cos(\theta^{\rm{az}}_{j})\\ 
\sin(\theta^{\rm{el}}_{j})
\end{bmatrix} , \rm{ with}  \\
d_{j}&\in [d_{\rm{min}}, d_{\rm{max}}] \; ,\; \theta^{\rm{az}}_{j}\in [\theta^{\rm{az}}_{\rm{min}}, \theta^{\rm{az}}_{\rm{max}}] \; , \notag \\
\theta^{\rm{el}}_{j}&\in [\sin^{-1}(\Delta z_{\rm{min}}/d_{j}), \sin^{-1}(\Delta z_{\rm{max}}/d_{j})] \quad \notag.
\end{align}
The selection of each focus point, $\bm{p}_{j}$ is not trivial and optimizing the design process is not explored further here. However, a large number of focus points need to be considered as the sweeping process needs to adequately cover $\mathcal{R}$ and thus the total number of spatial beams is dictated by the total size of this region.

\section{UE Localization}
In this section, we explore the practical localization algorithms, we study their complexity and highlight their purposes. Compared to the NF-based approach, the FF-based approach allows for a lower complexity localization algorithm with the relaxed channel model in~\eqref{eq:channel_h_FF}, at the cost of accuracy. The lower signaling overhead of the beam design also affects the computational load. With our proposed scheme, the high complexity algorithm is used \emph{only when required}.

\subsection{FF Localization Algorithm}
The process for the FF-based localization procedure is explored here, as summarized in Alg.~\ref{alg:pos_scheme_1}. Initially, there is a weighting procedure across the $J_{1}$ beams, with the normalized weight of each beam, $w_{j}$, acting as an RSSI. This encapsulates how much energy is received along each spatial beam of $\bm{F}_{1}$ and serves as a metric for the compatibility of $\theta_{j}$ in the beam design process and the actual AOD. Note that for a normalised RSSI below a threshold, $\varepsilon_{w}$, we disregard the corresponding beam, as it may lead to unreliable estimations. The position estimation consists of an iterative estimation procedure along the DOF and AOD. Each iteration of the process is split into three separate maximization searches, since the AOD is a two-dimensional variable. The relaxed channel model in~\eqref{eq:channel_h_FF} decouples the range and the angular domain.\footnote{In~\eqref{eq:channel_h_FF}, the DOF affects the phase offset along subcarriers only, while the AOD affects the phase offset among receiving elements only.} The DOF estimation relies on the phase offset between subcarriers and each beam is processed individually, resulting into $J_{1}$ DOF estimations. The RSSI indicator provides a weighted mean for each iteration. On the other hand, the AOD estimation relies on the phase offset between antenna elements and each subcarrier is processed individually. 

Since the range is decoupled from the angular domain, the iterative procedure will converge as long as each separate maximization search converges. Note that there is a maximum resolvable range, dictated by the subcarrier spacing\footnote{The non ambiguous range is given by $c_{\rm{o}}/W_{\rm{sub}}$.}, $W_{\rm{sub}}$. The computational complexity depends on the number of iterations $I_{1}$, the number of search steps in each maximization process, $I_{d}$ and $I_{\theta}$, the number of spatial beams, $J_{1}$, and the computational cost of reconstructing a FF steering vector and a subcarrier phase offset vector in~\eqref{eq:t}. See Section~\ref{sec:complexity} for a detailed discussion and comparison with the NF localization algorithm.  

\begin{algorithm}
\caption{Localization assuming FF model}\label{alg:pos_scheme_1} 
\begin{algorithmic}
\State Define $\varepsilon_{w} \in\mathbb{R}, I_{1} , I_{d} , I_{\theta} \in\mathbb{N}$
\State Input: $\bm{Y}, \bm{p}_{\rm{BS}}, \mathcal{R}$ 
\For{$j=1:J_{1}$}
    \State $w_{j} = \|\bm{y}_{j}\|/\text{max}(\|\bm{y}_{1:J_{1}}\|)$
    \If{$w_{j} < \varepsilon_{w}$}
        \State $w_{j} = 0$
    \EndIf
\EndFor
\State $\tilde{\bm{F}}_{1} = \bm{F}_{1}\rm{diag}(\bm{w})$
\For{$ii = 1:I_{1}$}
    \For{$j=1:J_{1}$} \Comment{$I_{d}$ search steps}
        \State $\hat{d}_{j} = \argmax_{d} \| \bm{y}\herm_{j}\bm{t}(d)\|$ 
    \EndFor
    \State $\hat{d} = \sum_{j=1}^{J_{1}}w_{j}\hat{d}_{j} / \sum_{j=1}^{J_{1}}w_{j}$
    \For{$q=1:Q$} \Comment{$I_{\theta}$ search steps}
        \State $\hat{\theta}^{\rm{az}}_{q} = \argmax_{\theta^{\rm{az}}} \| \bm{\alpha}(\bm{\theta})\herm\tilde{\bm{F}}_{1}\bm{y}_{q} \| : \bm{\theta}=[\theta^{\rm{az}}, \hat{\theta}^{\rm{el}}]$ 
    \EndFor
    \State $\hat{\theta}^{\rm{az}} = \sum_{q=1}^{Q} \hat{\theta}^{\rm{az}}_{q}/ Q$
    \For{$q=1:Q$} \Comment{$I_{\theta}$ search steps}
        \State $\hat{\theta}^{\rm{el}}_{q} = \argmax_{\theta^{\rm{el}}} \| \bm{\alpha}(\bm{\theta})\herm\tilde{\bm{F}}_{1}\bm{y}_{q} \| : \bm{\theta}=[\hat{\theta}^{\rm{az}}, \theta^{\rm{el}}]$ 
    \EndFor
    \State $\hat{\theta}^{\rm{az}} = \sum_{q=1}^{Q} \hat{\theta}^{\rm{az}}_{q}/ Q$

\EndFor
\State $\hat{\bm{p}}_{\rm{UE}} = \bm{p}_{\rm{BS}} + \hat{d} \begin{bmatrix}
\cos(\hat{\theta}^{\rm{el}})\sin(\hat{\theta}^{\rm{az}})\\ 
\cos(\hat{\theta}^{\rm{el}})\cos(\hat{\theta}^{\rm{az}})\\ 
\sin(\hat{\theta}^{\rm{el}})
\end{bmatrix}$ \Comment{Output}
\end{algorithmic}
\end{algorithm}

\subsection{NF Localization Algorithm}
The process for the NF-based localization procedure is explored here, as summarized in Alg.~\ref{alg:pos_scheme_2}. The position estimation is iterative maximization along the DOF and AOD. Since the range and angular domains are not decoupled, a starting point is not trivial to derive. To avoid multiple iterative maximization procedures with different starting points, we choose the focus point of the spatial stream of $\bm{F}_{2}$ with the highest RSSI weight coefficient, $w_{j}$. Similar to Alg.~\ref{alg:pos_scheme_1}, the weighting coefficient of beams below a thresehold is set to 0, as their contribution may be problematic. In order for this process to converge, the focus-points of $\bm{F}_{2}$ need to be dense enough to ensure proper coverage of $\mathcal{R}$, which requires a larger $J_{2}$ and thus a higher signaling overhead. 

A higher signaling overhead not only requires additional power and time resources, but also increases the computational complexity of the algorithm. In Alg.~\ref{alg:pos_scheme_2} each maximization search is performed across each subcarrier separately and the final parameter estimation takes the mean across all subcarriers. Note that each search step in Alg.~\ref{alg:pos_scheme_2} requires the reconstruction of the NF steering vector, $\bm{\alpha(\bm{p})}$, for both the angular and the range domain. Other than $J_{2}$, the computational complexity depends on the number of iterations $I_{2}$ and the computational cost of reconstructing a NF steering vector in~\eqref{eq:sv_NF}.  

\begin{algorithm}
\caption{Localization assuming NF model}\label{alg:pos_scheme_2} 
\begin{algorithmic}
\State Define $\varepsilon_{w} \in\mathbb{R}, I_{2} , I_{d} , I_{\theta} \in\mathbb{N}$
\State Input: $\bm{Y}, \bm{p}_{\rm{BS}}, \mathcal{R}$ 
\For{$j=1:J_{2}$}
    \State $w_{j} = \|\bm{y}_{j}\|/\text{max}(\|\bm{y}_{1:J_{2}}\|)$
    \If{$w_{j} < \varepsilon_{w}$}
        \State $w_{j} = 0$
    \EndIf
\EndFor
\State $j\ast = \argmax_{j} w_{j} $
\State $\hat{\bm{p}} = \bm{p}_{j\ast}$ \Comment{focus point of $\bm{f}_{2,j\ast}$}
\State $\tilde{\bm{F}}_{2} = \bm{F}_{2}\rm{diag}(\bm{w})$
\For{$ii = 1:I_{2}$}
    \For{$q=1:Q$} \Comment{$I_{d}$ search steps}
        \State $\hat{d}_{q} = \argmax_{d} \| \bm{\alpha}(\bm{p})\herm\tilde{\bm{F}}_{2}\bm{y}_{q} \|$ 
    \EndFor
    \State $\hat{d} = \sum_{q=1}^{Q}\hat{d}_{q} / Q$
    \For{$q=1:Q$} \Comment{$I_{\theta}$ search steps}
        \State $\hat{\theta}^{\rm{az}}_{q} = \argmax_{\theta^{\rm{az}}} \| \bm{\alpha}(\bm{p})\herm\tilde{\bm{F}}_{2}\bm{y}_{q} \|$ 
    \EndFor
    \State $\hat{\theta}^{\rm{az}} = \sum_{q=1}^{Q} \hat{\theta}^{\rm{az}}_{q}/ Q$
    \For{$q=1:Q$} \Comment{$I_{\theta}$ search steps}
        \State $\hat{\theta}^{\rm{el}}_{q} = \argmax_{\theta^{\rm{el}}} \| \bm{\alpha}(\bm{p})\herm\tilde{\bm{F}}_{2}\bm{y}_{q} \|$ 
    \EndFor
    \State $\hat{\theta}^{\rm{az}} = \sum_{q=1}^{Q} \hat{\theta}^{\rm{az}}_{q}/ Q$

\EndFor
\State $\hat{\bm{p}}_{\rm{UE}} = \bm{p}_{\rm{BS}} + \hat{d} \begin{bmatrix}
\cos(\hat{\theta}^{\rm{el}})\sin(\hat{\theta}^{\rm{az}})\\ 
\cos(\hat{\theta}^{\rm{el}})\cos(\hat{\theta}^{\rm{az}})\\ 
\sin(\hat{\theta}^{\rm{el}})
\end{bmatrix}$ \Comment{Output}
\end{algorithmic}
\end{algorithm}

\subsection{Complexity Analysis}\label{sec:complexity}
The dominant computational load in both cases is the alternating maximization and the 3-dimensional grid search, which is essentially a non-convex optimization problem. The resolution of the grid search and the number of iterations depend on the desired accuracy and the complexities of both algorithms grow linearly with these parameters. The relevant computational complexities are summarised in Table~\ref{tab:complexity}. 

The size of the BS is shown to affect the overall complexity the most, while the signaling overhead, i.e., the number of transmitted beams also affects the complexity, giving a computational edge to Alg.~\ref{alg:pos_scheme_1}. Other than $J_{2}$ being larger than $J_{1}$, the distinctive computational disadvantage of Alg.~\ref{alg:pos_scheme_2} is that each grid search requires the reconstruction of $\bm{\alpha}$, which depends an all three parameters that need to be estimated. Without a simplified closed-form expression for each vector element, the NF steering vector$\bm{\alpha}$ is notably more complex than the FF $\tilde{\bm{\alpha}}$. 

\begin{figure}[h]
    \centering
    \includegraphics[width=\columnwidth]{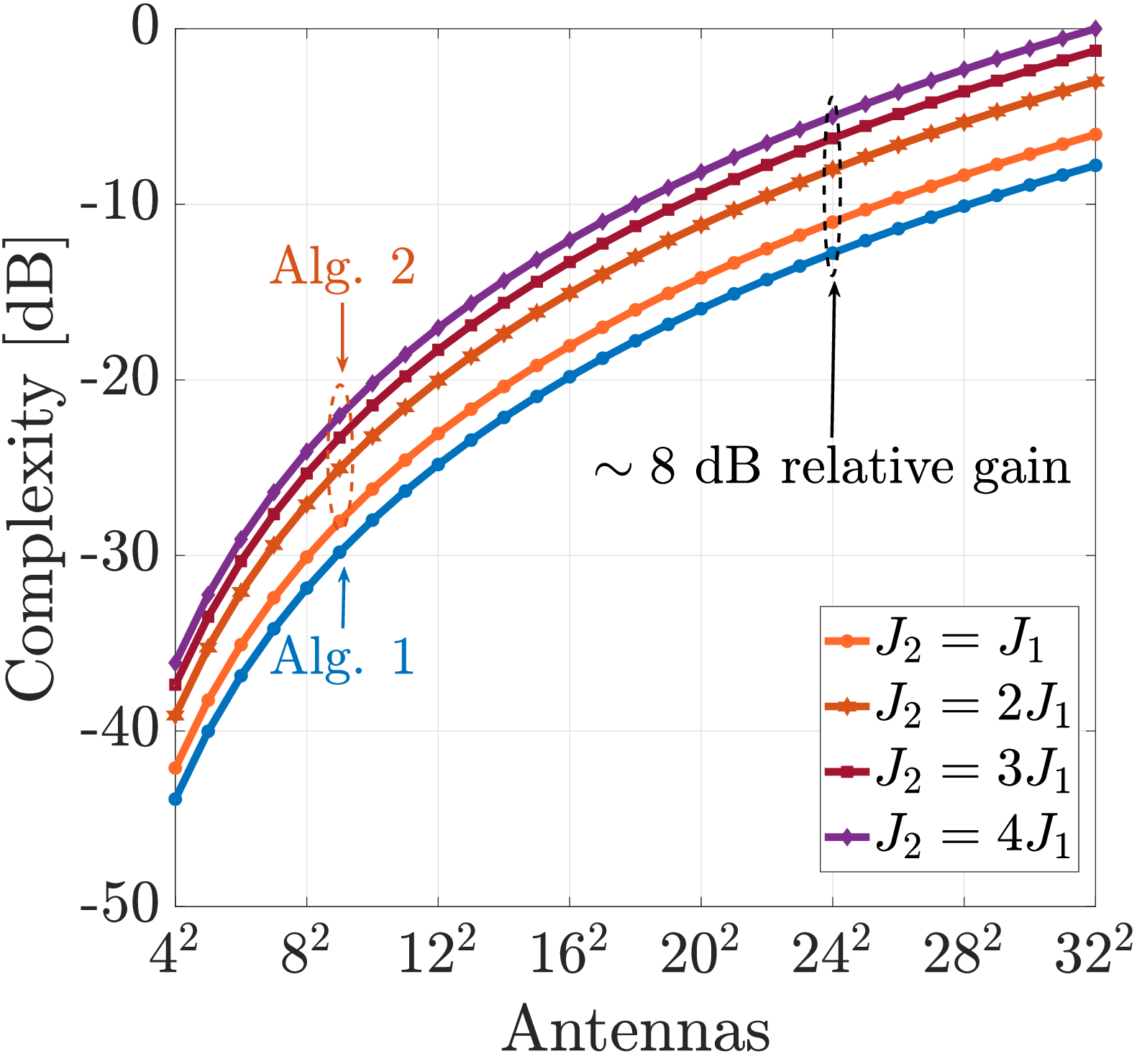}
    \caption{Normalised relative complexity for different signaling overhead requirements for the two proposed Algorithms.}
    \label{fig:comp}
\end{figure}

Numerical evaluation highlights the computational mismatch of the two signaling schemes. Figure~\ref{fig:comp} shows the complexity growth as the BS array becomes larger. Different signaling overhead requirements are shown for the NF-based scheme, relative to the FF-based scheme. We observe that the complexity rises exponentially as the number of antennas increases. When the FF-based scheme is employed, a relative computational gain is observed compared to the NF-based scheme, even when $J_{1} = J_{2}$. However, the NF-based scheme is expected to have higher signaling overhead. When Alg.~\ref{alg:pos_scheme_2} has 300\% larger signaling overhead, we observe approximately 8 dB of computational gain, i.e., when $J_{2}=4J_{1}$ we observe a run time 6.3 times longer for Alg.~\ref{alg:pos_scheme_2}, compared to Alg.~\ref{alg:pos_scheme_1}. The values here are normalized relative to the largest value. Overall, a definitive closed-form expression can not be derived and it needs to be noted that both algorithms do not produce optimal estimations, but rather they are designed with a similar practical structure that highlights the limitation of beam-steering generalizations and relaxed channel assumptions in localization applications. 

\begin{table}
\caption{Complexity analysis of Algorithms~\ref{alg:pos_scheme_1} \& \ref{alg:pos_scheme_2}.}
\label{tab:complexity}
\begin{center}
\renewcommand{\arraystretch}{1.5} 
\begin{tabular}{l l}
    \hline
    \multicolumn{2}{c}{\cellcolor{lightgray}{Complexity Analysis}} \\
    \hline	
    $\mathcal{O}\{ \bm{t}(d)\}$ & $\propto Q$ : One set of multiplications per subcarrier. \\
    \hline
    $\mathcal{O}\{ \tilde{\bm{\alpha}}(\bm{\theta})\}$ & $\propto N_{\rm{BS}}$ : One set of multiplications per antenna element.\\ 
    \hline
    $\mathcal{O}\{ \bm{\alpha}(\bm{p})\}$ & $\propto N_{\rm{BS}} \cdot\mathcal{O}\{ \|\bm{u} - \bm{v} \|\}$ : One set of multiplications and \\ &\quad an exact distance to be derived per antenna element.\\
    \hline
    $\mathcal{O}\{ \rm{Alg.~\ref{alg:pos_scheme_1}} \}$ & $\propto I_{1}I_{d}J_{1}\bigl(\mathcal{O}\{ \bm{t}(d)\} + Q\bigr) +$\\&\quad$ 2I_{1}I_{\theta}Q\bigl(\mathcal{O}\{ \tilde{\bm{\alpha}}(\bm{\theta})\} + N_{\rm{BS}}^{2}J_{1}\bigr)$ \\
    \hline
    $\mathcal{O}\{ \rm{Alg.~\ref{alg:pos_scheme_2}} \}$ & $\propto I_{1}Q(I_{d}+2I_{\theta})\bigl(\mathcal{O}\{ \bm{\alpha}(\bm{p})\}+ N_{\rm{BS}}^{2}J_{2}\bigr)$ \\
    \hline
\end{tabular}
\end{center}
\end{table}

\section{UE Tracking \& Adaptive Positioning}\label{sec:track}
In this section, we introduce a tracking application that uses a repetitive localization procedure. We highlight various degrees of freedom that such an application provides and investigate the effect of different design parameters in the overall behaviour of the system. The procedure is scalable to the cases with multiple UEs, since the BS sweeps the whole region and each UE estimates its own position. However, for ease of notation and to avoid a complicated adaptive signaling procedure, a single UE is considered in the following. We consider multiple iterations, say $K$, with the UE moving in-between iterations. As the UE moves, it collects positional information and forms its trajectory, which introduces a sense of memory in the system. The structure of the tracking procedure is described in Alg.~\ref{alg:track_proc}, where the core ideas are:
\begin{itemize}
\item \emph{Initial Signaling Decision}: For each iteration, the BS needs to decide which signaling scheme needs to be employed. The binary scalar $\xi_{k}\in\{1,2\}$ encodes the selected scheme. The initial choice depends, among other factors, on the desired positioning accuracy, the expected UE position based on the formed trajectory and if previous estimations are stored in the memory buffer, $\mathcal{M}$. When no prior information is available, the reliability of the initial estimation is crucial to be properly validated. 
\item \emph{Forming a trajectory}: For each iteration, the previously derived position estimations can be utilized to form a trajectory and improve the overall system performance. A low-complexity polynomial fitting algorithm may be employed for each dimension. The polynomial degree, $n_{\rm{d}}$, encapsulates the UE movement model. To form such a trajectory, a minimum number of $n_{\rm{d}}+1$ estimations are required, while $K_{\rm{F}}$ serves as a cut off point for older estimations. Note that the tracking model and the trajectory formation can be improved, by employing a weighted polynomial formation or a Kalman filter.
\item \emph{Additional Signaling}: For each iteration, the BS needs to decide whether an additional signaling scheme is necessary, when the position estimation is unreliable. The estimated DOF and the formed trajectory serve as reliability metrics. When the estimated position is below the NF distance, $d_{\rm{NF}}$, then the FF-based estimation is considered unreliable. Note that $d_{\rm{NF}}$ is not necessarily the FD, or any other general indicator of the NF region, but rather it refers to the range-limit of implementing the FF-based scheme reliably, for this specific geometry and system parameters. A more thorough discussion about the NF-FF limit is presented in Sections~\ref{sec:FF}\&~\ref{sec:sim}. In addition, an estimation is deemed unreliable when it diverges from the expected position. For an acceptable estimation, the trajectory-estimation error $\mathcal{E}_{\rm{UE}}$ may not exceed an error threshold of $\varepsilon_{\rm{th}}$. As the two signaling schemes have different positioning accuracy levels, previous estimations may be incompatible when the signaling scheme changes and the memory may need to be truncated. Note that the switching protocol may involve a more complicated procedure that takes into account the increased complexity of the NF-based scheme and the computational cost of switching the signaling protocol.
\end{itemize}
There are various tracking techniques, each with numerous degrees of freedom. Our proposed tracking algorithm serves to highlight the signaling adaptability of a system when a sense of memory is introduced. The adaptive nature of such a system improves the signaling and computational complexity of the positioning application.
\begin{algorithm}
\caption{Tracking Process}\label{alg:track_proc} 
\begin{algorithmic}
\State Initialize: $K_{\rm{F}}\in\mathbb{N},n_{d}\in\mathbb{N},\xi_{1}\in\{1,2\},\varepsilon_{\rm{th}}\in\mathbb{R},d_{\rm{NF}}\in\mathbb{R}$.
\State Initialize empty memory: $\mathcal{M} = \{\}$.
\State Note: $\acute{\bm{p}}_{\rm{UE}}$ refers to the estimated UE position based on the formed trajectory.
\For{$k=1:K$}
    \If{$\xi_{k} = 1$}
        \State Employ \emph{Signaling Scheme} 1 for $\hat{\bm{p}}_{\rm{UE}}^{(k)}$.
        \If{$\rm{card}\{\mathcal{M}\} > n_{d}+1$}
            \State Load up to $K_{\rm{F}}$ latest inputs from $\mathcal{M}$.
            \State $\acute{\bm{p}}_{\rm{UE}}^{(k)} = $ Polynomial fitting of $n_{d}$ degree.
        \Else
            \State $\acute{\bm{p}}_{\rm{UE}}^{(k)} = \hat{\bm{p}}_{\rm{UE}}^{(k)}$. 
        \EndIf
        \State Trajectory-Estimation error: $\mathcal{E}_{\rm{UE}}^{(k)} = \|\hat{\bm{p}}_{\rm{UE}}^{(k)} - \acute{\bm{p}}_{\rm{UE}}^{(k)} \|$.
        \If{$\mathcal{E}_{\rm{UE}}^{(k)} > \varepsilon_{\rm{th}}$ or $\|\hat{\bm{p}}_{\rm{UE}}^{(k)}\| < d_{\rm{NF}}$}
            \State Empty memory: $\mathcal{M} = \{\}$. \Comment{Optional}
            \State Employ \emph{Signaling Scheme} 2 for $\hat{\bm{p}}_{\rm{UE}}^{(k)}$. 
            \State Store $\hat{\bm{p}}_{\rm{UE}}^{(k)}$ to memory: $\mathcal{M}$.
        \Else
            \State Update trajectory $\acute{\bm{p}}_{\rm{UE}}^{(k)}$ including $\hat{\bm{p}}_{\rm{UE}}^{(k)}$.
            \State Store $\acute{\bm{p}}_{\rm{UE}}^{(k)}$ to memory: $\mathcal{M}$.
        \EndIf
    \ElsIf{$\xi_{k} = 2$}
        \State Employ \emph{Signaling Scheme} 2 for $\hat{\bm{p}}_{\rm{UE}}^{(k)}$.
        \If{$\rm{card}\{\mathcal{M}\} > n_{d}$}
            \State Load $\hat{\bm{p}}_{\rm{UE}}^{(k)}$ and up to $K_{\rm{F}}$ latest inputs from $\mathcal{M}$.
            \State $\acute{\bm{p}}_{\rm{UE}}^{(k)} = $ Polynomial fitting of $n_{d}$ degree.
            \State Store $\acute{\bm{p}}_{\rm{UE}}^{(k)}$ to memory: $\mathcal{M}$.
        \Else
            \State Store $\hat{\bm{p}}_{\rm{UE}}^{(k)}$ to memory: $\mathcal{M}$.
        \EndIf
    \EndIf
    \If{$\hat{\bm{p}}_{\rm{UE}}^{(k)} \in\rm{FF}$ and $k<K$} \Comment{Next Iteration}
        \State $\xi_{k+1} = 1$
    \ElsIf {$\|\hat{\bm{p}}_{\rm{UE}}^{(k)}\| < d_{\rm{NF}}$ and $k<K$}
        \State $\xi_{k+1} = 2$
    \EndIf
\EndFor
\end{algorithmic}
\end{algorithm}

\section{Far-Field Limitation}\label{sec:FF}
Here, we explore the limitations of the FF-based scheme in terms of appropriate beam design and positioning accuracy. Traditionally, the NF-FF border solely relates to the spherical nature of the wavefront. However, when the whole localization scheme is considered, a large BS-to-UE distance allows for lower complexity design of beams and positioning algorithms, which are not necessarily tied to the standard interpretations of FF. It is convenient, when the geometry allows it, to consider the DOF as the deciding factor for the performance of such a scheme. To further explore this idea, we observe the noise free signal
\begin{align}
\tilde{\bm{s}}\transp(\bm{p}) &= \beta\sqrt{\mathcal{P}}\bm{f}_{\rm{o}}\herm\bm{A}(\bm{p}) \;, 
\end{align}
where all OFDM symbols are assumed to be 1 and only a single beam is considered to avoid complex notation. Here, 
\begin{align}
\bm{A}(\bm{p}) &= [\bm{\alpha}_{1}(\bm{p}) \cdots \bm{\alpha}_{q}(\bm{p}) \cdots \bm{\alpha}_{Q}(\bm{p}) ] \; ,
\end{align}
is the collection of steering vectors across all subcarriers. We can disregard the irrelevant scaling factor of $\beta\sqrt{\mathcal{P}}$ and have
\begin{align}
\bm{s}\transp(\bm{p}) &= \bm{f}_{\rm{o}}\herm\bm{A}(\bm{p}) \;. 
\label{eq:s}
\end{align}
As discussed in Section~\ref{ssec:beam}, for the FF-based approach $\bm{f}$ is a FF-steering vector. Without loss of generality, we can choose $\bm{f}_{\rm{o}} = \tilde{\bm{\alpha}}([0,0]) = \mathbf{1}$ and the $q^{\rm{th}}$ element of $\bm{s}(\bm{p})$ is
\begin{align}
s_{q} = \sum_{n=1}^{N_{\rm{BS}}}\frac{d\lambda_{q}}{d_{n}\lambda_{\rm{o}}} \text{exp}\bigl\{-\j\frac{2\pi}{\lambda_{q}}d_{n} \bigr\} \; ,
\end{align}
where $d = \|\bm{p}_{\rm{BS}}-\bm{p} \|$ and $d_{n} = \|\bm{p}^{(n)}_{\rm{BS}}-\bm{p} \|$. Note that the amplitude of $s_{q}$ encapsulates whether the transmitting elements interfere constructively at $\bm{p}$, when the beam was designed under the FF assumption. 

We introduce the normalised scalar metric
\begin{align}
\label{eq:g}
g(\bm{p}) &= \frac{1}{N_{\rm{BS}}Q}|\bm{s}\herm(\bm{p})\bm{t}(d)| , \\
g(\bm{p}) &= \frac{1}{N_{\rm{BS}}Q}  \notag \\ & \quad\times \biggl| \sum_{q=1}^{Q}\sum_{n=1}^{N_{\rm{BS}}}\frac{d\lambda_{q}}{d_{n}\lambda_{\rm{o}}} \text{exp}\bigl\{-\j2\pi\bigl[(q-1)\frac{W_{\rm{sub}}}{c_{\rm{o}}}d - \frac{d_{n}}{\lambda_{q}} \bigr]\bigr\}  \biggr|\; \notag,
\end{align}
which combines the successive subcarriers of~\eqref{eq:s} with the rolling phase between successive subcarriers of~\eqref{eq:t}; an essential component of the relaxed FF model in~\eqref{eq:channel_h_FF}. Essentially~\eqref{eq:g} is a normalised compatibility indicator for the FF-beam design and the range-angular domain detachment utilized in Alg.~\ref{alg:pos_scheme_1}. We observe that $g(\bm{p})$ is the amplitude of a sum of complex scalars and it is maximized when the summed elements are in phase, which is true when $d_{n}=d,\;\forall n = 1,\dots,N_{\rm{BS}}$. In other words, variance in the DOF for different antenna elements results in larger phase variance within the complex sum and thus a lower compatibility. The variance in antenna-to-receiver distances depends on the considered point-receiver, $\bm{p}$, and the compatibility metric is deterministic for a given point. However, when we consider a fixed BS-to-UE distance the metric is a random variable, as there are infinite points in $\mathcal{R}$ with that given distance. Region $\mathcal{R}$ is structured in way where a sweep across the azimuth domain is always required. 

\begin{figure}[h]
    \centering
    \includegraphics[width=\columnwidth]{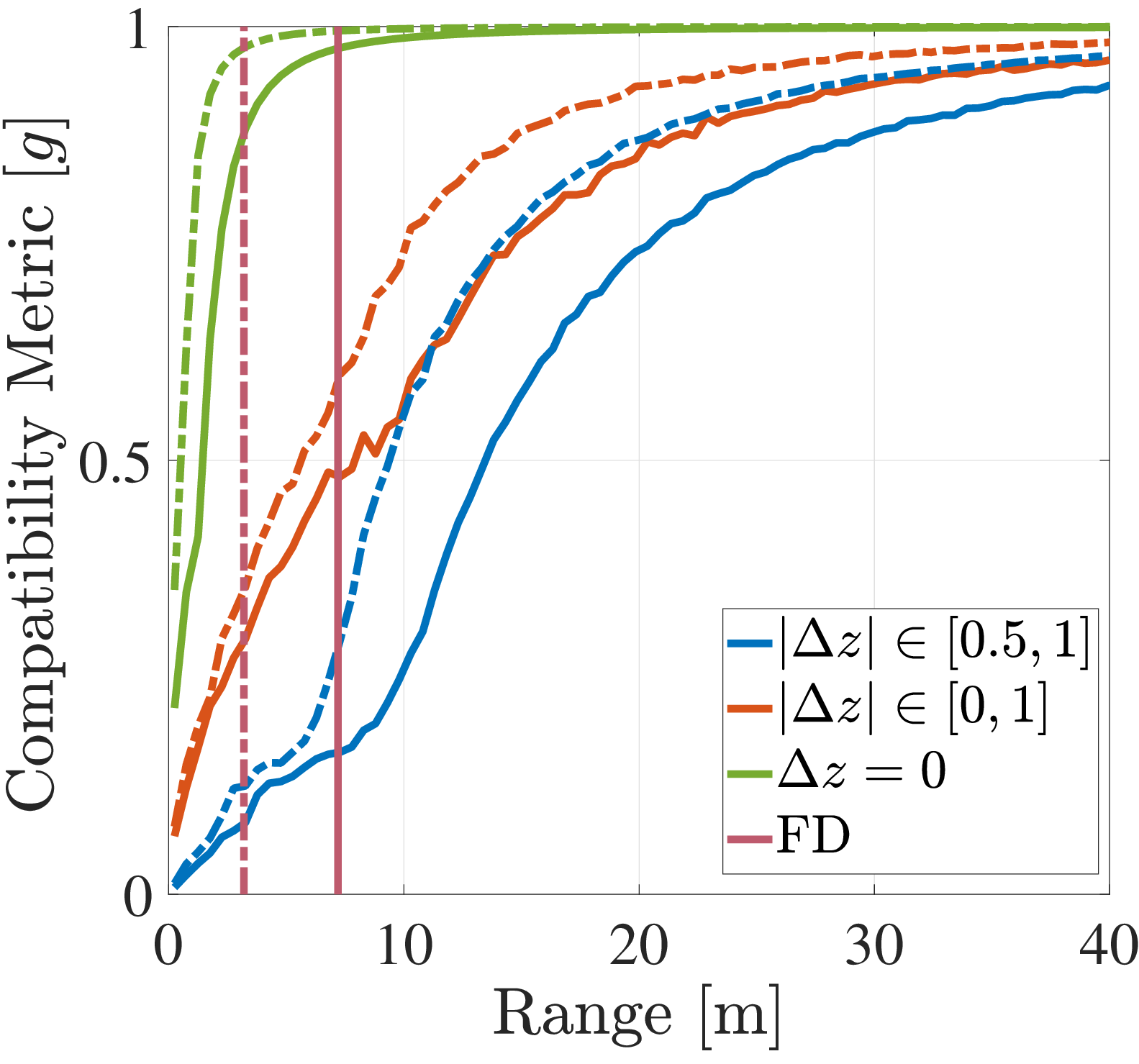}
    \caption{Normalised compatibility metric for different UE distributions and different array sizes. Solid lines refer to $N_{\rm{BS}}=24\times 24$ and dashed lines to $N_{\rm{BS}}=16\times 16$.}
    \label{fig:g_metric}
\end{figure}

Since~\eqref{eq:g} considers a single beam, we disregard the azimuth mismatch, which is addressed by the beam-sweep. Therefore, we examine the compatibility of a receiver in $\mathcal{R}$ as the DOF increases, while the azimuth component is perfectly matched to the transmitted beam, i.e., $\theta^{\rm{az}}$. The only variable that may introduce a beam mismatch, other than the DOF, is the height mismatch between the BS and the receiver, $\Delta z$. Figure~\ref{fig:g_metric} shows that a larger height mismatch decreases the compatibility of the signaling scheme and the positioning algorithm that are designed under the assumption of a large BS-to-UE distance. The compatibility of that design approaches unity as the range increases. Moreover, even when the FF-designed beam perfectly matches the UE ($\Delta z = 0$), the compatibility metric always decreases within the standard NF region, defined by the FD.  

\section{Simulation Results}\label{sec:sim}
In this section, we present simulation evaluations and different simulation setups that explore the performance of the two proposed schemes, highlight the complexity-accuracy trade-off and validate the significance of a repetitive tracking procedure. Consider Fig.~\ref{fig:sm} and the simulation parameters given in Table~\ref{tab:sim_par}. We consider different scenarios with different BS array sizes. Unless otherwise specified, Scenario A is employed. However, we have tested all results for different scenarios and observed the same qualitative conclusions as those presented in the figures.
\begin{table}
\caption{Simulation Parameters}
\label{tab:sim_par}
\begin{center}
\renewcommand{\arraystretch}{1.5} 
\begin{subtable}[h]{\textwidth}
\begin{tabular}{l  l | l  l | l l}
    \hline
    \multicolumn{6}{c}{\cellcolor{lightgray}{Simulation Parameters}} \\
    \hline	
    $J_{1}$      & 21        & $J_{2}$               &    84                 & $\bm{p}_{\rm{BS}}$             & [0,0, 2]             \\
    $Q$          & 12        & $W_{\rm{o}}$          & 15 KHz                & $W_{\rm{sub}}$                 & 750 KHz              \\
    BW           & 8.265 MHz & $\delta_{\rm{BS}}$    & $\lambda_{\rm{o}} /2$ & $\theta_{\rm{UE}}^{(\rm{az})}$ & $\in [-\pi/4,\pi/4]$ \\
    $f_{\rm{o}}$ & 24 GHz    & $\rm{Noise}_{\rm{F}}$ & 10 dB                 & $\rm{Noise}_{\rm{D}}$          & -174 dBm/Hz          \\
    \hline
\end{tabular}
\end{subtable}
\begin{subtable}[h]{\textwidth}
\begin{tabular}{l | l  l | l  l | l l | l l }

    \multicolumn{9}{c}{\cellcolor{lightgray}{Different Scenarios}} \\
    \hline	
    A$\;\;$ & \; $N_{\rm{BS}}^{(\rm{x})}$ & 24 & $N_{\rm{BS}}^{(\rm{z})}$ & 24 & FD & 7.2 m & $z_{\rm{UE}}$ & $\in [1,1.5]$ \\
    B       & \; $N_{\rm{BS}}^{(\rm{x})}$ & 16 & $N_{\rm{BS}}^{(\rm{z})}$ & 16 & FD & 3.2 m & $z_{\rm{UE}}$ & $\in [1,1.5]$ \\
    C       & \; $N_{\rm{BS}}^{(\rm{x})}$ & 16 & $N_{\rm{BS}}^{(\rm{z})}$ & 8  & FD & 3.2 m & $z_{\rm{UE}}$ & $\in [1,2]$ \\
    \hline
\end{tabular}
\end{subtable}
\end{center}
\end{table}

\begin{figure}[h]
    \centering
    \includegraphics[width=\columnwidth]{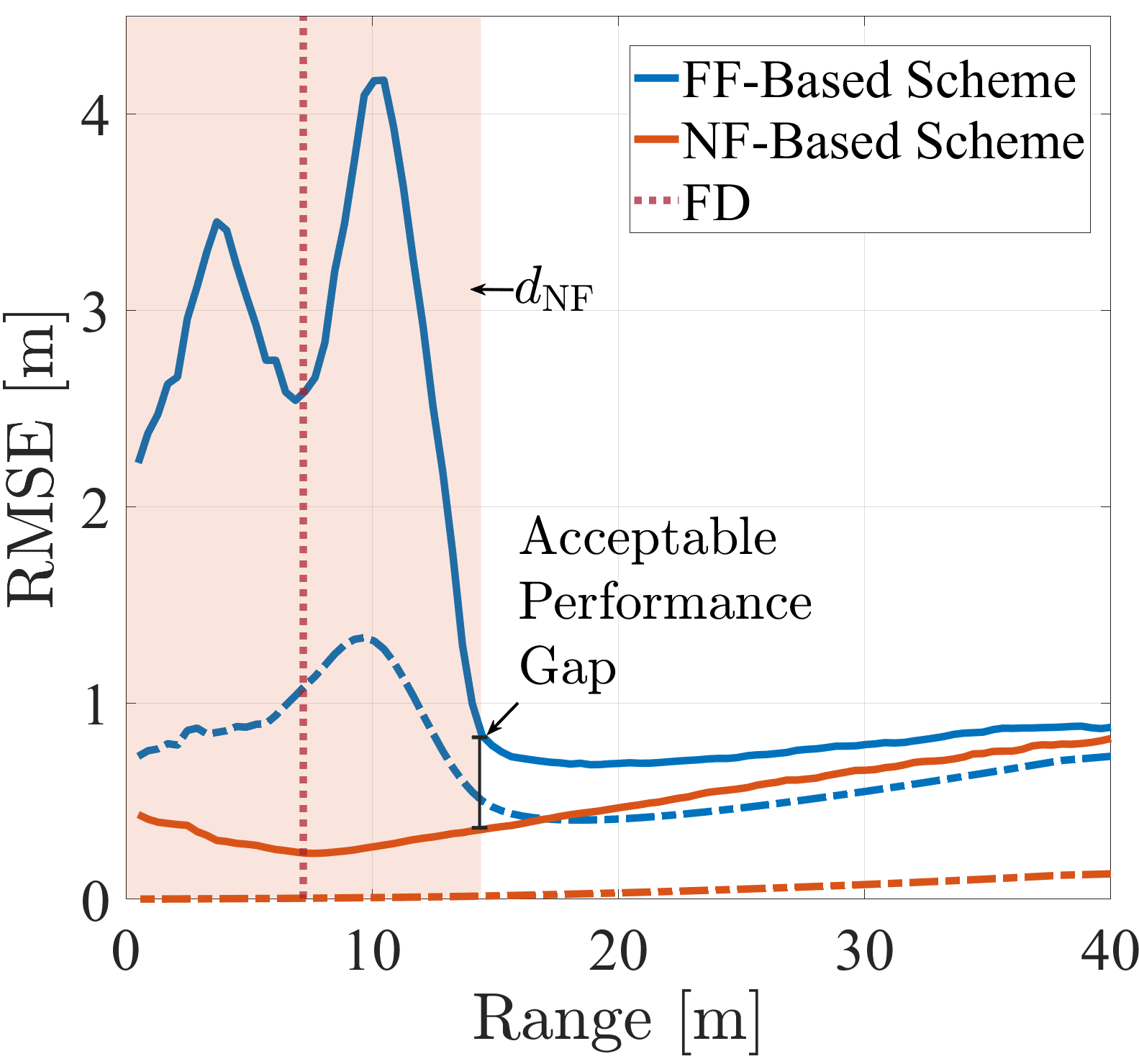}
    \caption{Position RMSE for the two proposed signaling schemes. The shaded region marks the possible $d_{\rm{NF}}$. The dashed lines mark the corresponding PEBs. Scenario A is considered.}
    \label{fig:pos_rmse}
\end{figure}

Figure~\ref{fig:pos_rmse} shows the positioning root mean square error (RMSE) in meters, for the two signaling schemes, versus the BS-to-UE distance. As expected, the NF-based signaling scheme outperforms the FF-based scheme, since more resources are utilized, i.e., computational power and transmitted pilots. While the signaling requirements and computational complexity are not affected by the BS-to-UE distance for both signaling setups, the positioning accuracy of the FF-based scheme is noticeably lower for UEs close to the BS. The FD distance is marked to highlight that it is not a good metric for the compatibility of the FF-based signaling design and the corresponding positioning algorithm. 

It needs to be noted that the NF-based scheme has a signaling overhead 400\% larger than the FF-based scheme, while the run-time of Alg.~\ref{alg:pos_scheme_2} is $\sim$8 times (9 dB) higher than Alg.~\ref{alg:pos_scheme_1}, which is in line with the complexity analysis in Table~\ref{tab:complexity}. In a practical scenario, the positioning accuracy needs to satisfy a performance thresehold, while the computational and signaling cost is kept low. In Fig.~\ref{fig:pos_rmse}, the shaded region marks the range that the NF-based scheme \emph{needs to be employed}. Therefore, it is shown that either of the two schemes should be employed, depending on the desired accuracy and the UE's position relative to the BS. For comparison, the position error bound (PEB) is shown when the true channel model in~\eqref{eq:channel_h_NF} and the relaxed channel model in~\eqref{eq:channel_h_FF} are used. The error bound refers to the Cramér Rao bound and marks the accuracy constraint for any estimator utilizing either channel model~\cite{chen2022channel}. It is evident that more positional information may be extracted from the true channel model from UEs closer to the BS, as the NF-PEB is significantly lower. Note that none of the NF- and FF-based schemes approach the optimal performance, as this would require, among other modifications, an exhaustive search of high complexity. Moreover, since the structure of the two algorithms is similar and the additional signaling of the NF-based scheme is focused close to the BS, their performance converges as the distance increases.

\begin{figure}[h]
    \centering
    \includegraphics[width=\columnwidth]{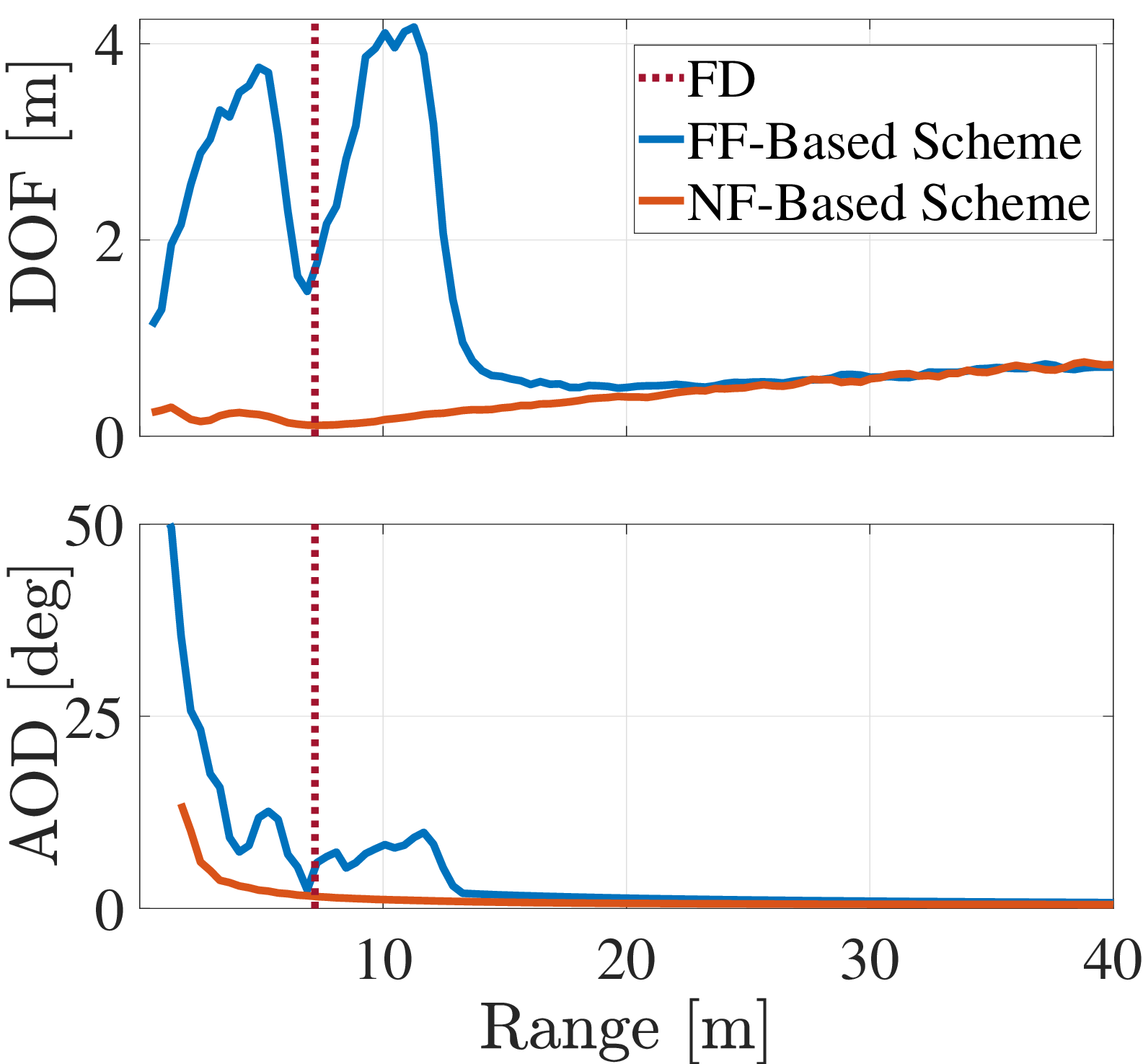}
    \caption{RMSE for the estimated parameters (DOF \& AOD) for the two proposed signaling scheme. Scenario A is considered.}
    \label{fig:dof_aod_rmse}
\end{figure}

Figure~\ref{fig:dof_aod_rmse} shows the RMSE for the range and angular domain, for the two schemes. The DOF estimation is highly unreliable for the FF-based scheme when the UE is close to the BS. The unreliable range estimation indicates that setting a hard boarder line between the two schemes is problematic, since this approach utilizes an unreliable metric, $\hat{d}$, for the evaluation of the estimation. In other words, the additional signaling decision, discussed in Section~\ref{sec:track} needs to rely more  on the formed trajectory, rather than comparing $\hat{d}$ itself with a constant $d_{\rm{NF}}$. In addition the AOD RMSE drastically increases when the UE is below the FD distance. The FF-based localization scheme may be incompatible due to the inadequate beam design or the spherical nature of the wavefront, and different estimation errors indicate that these factors may affect different estimated parameters to different extends. The increased $d_{\rm{NF}}$, relative to the FD, in Figs.~\ref{fig:pos_rmse} and~\ref{fig:dof_aod_rmse} appears analogous to the compatibility metric, $g$, dropping below 50\% in Fig.~\ref{fig:g_metric}. 

\begin{figure}[h]
    \centering
     \begin{subfigure}[t]{\columnwidth}
         \includegraphics[width=\columnwidth]{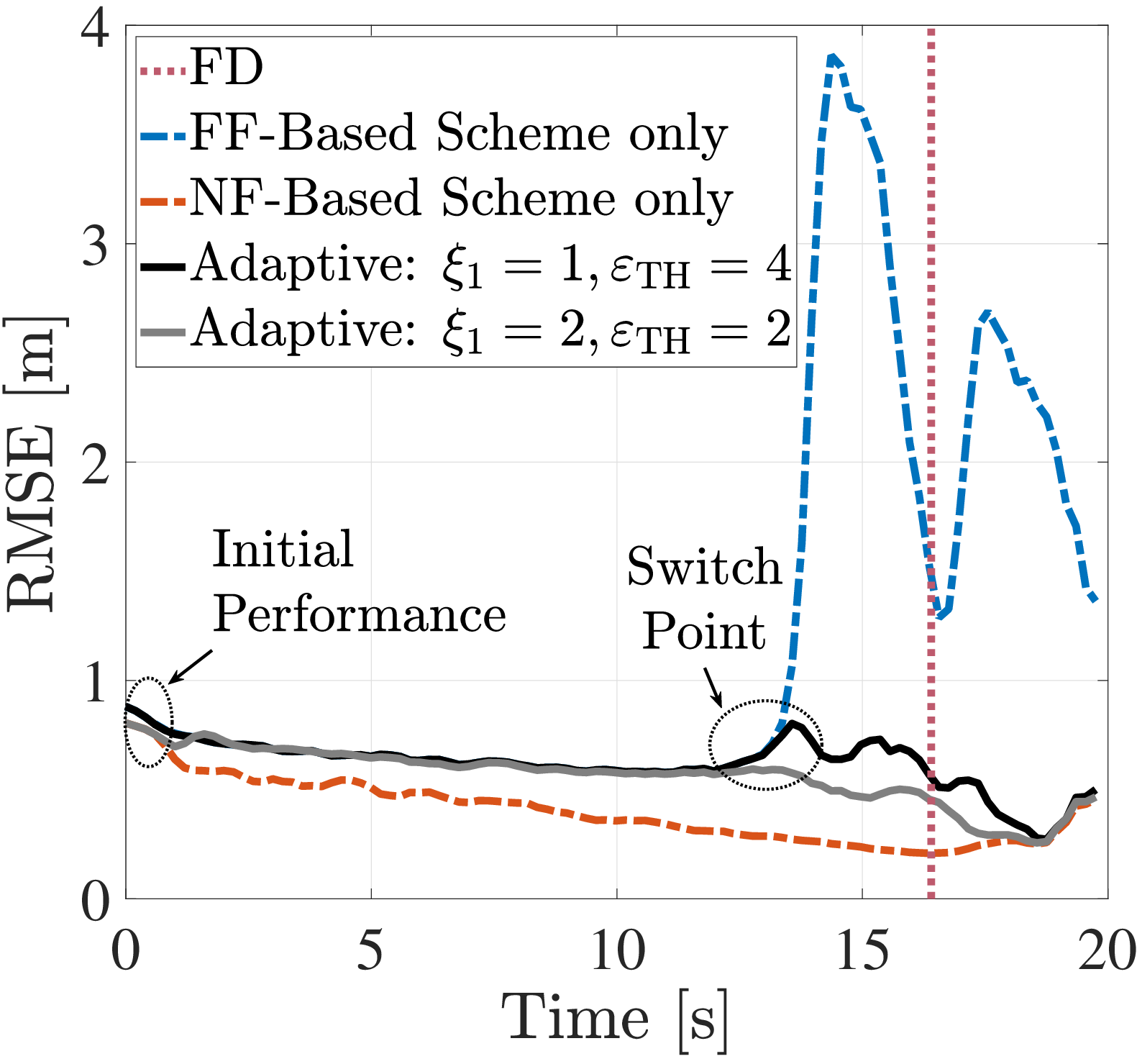}
         \caption{UE moving towards the BS at 2 m/s.}
         \label{sfig:track_rmse_a}
     \end{subfigure}
     \hfill
     \begin{subfigure}[t]{\columnwidth}
         \includegraphics[width=\columnwidth]{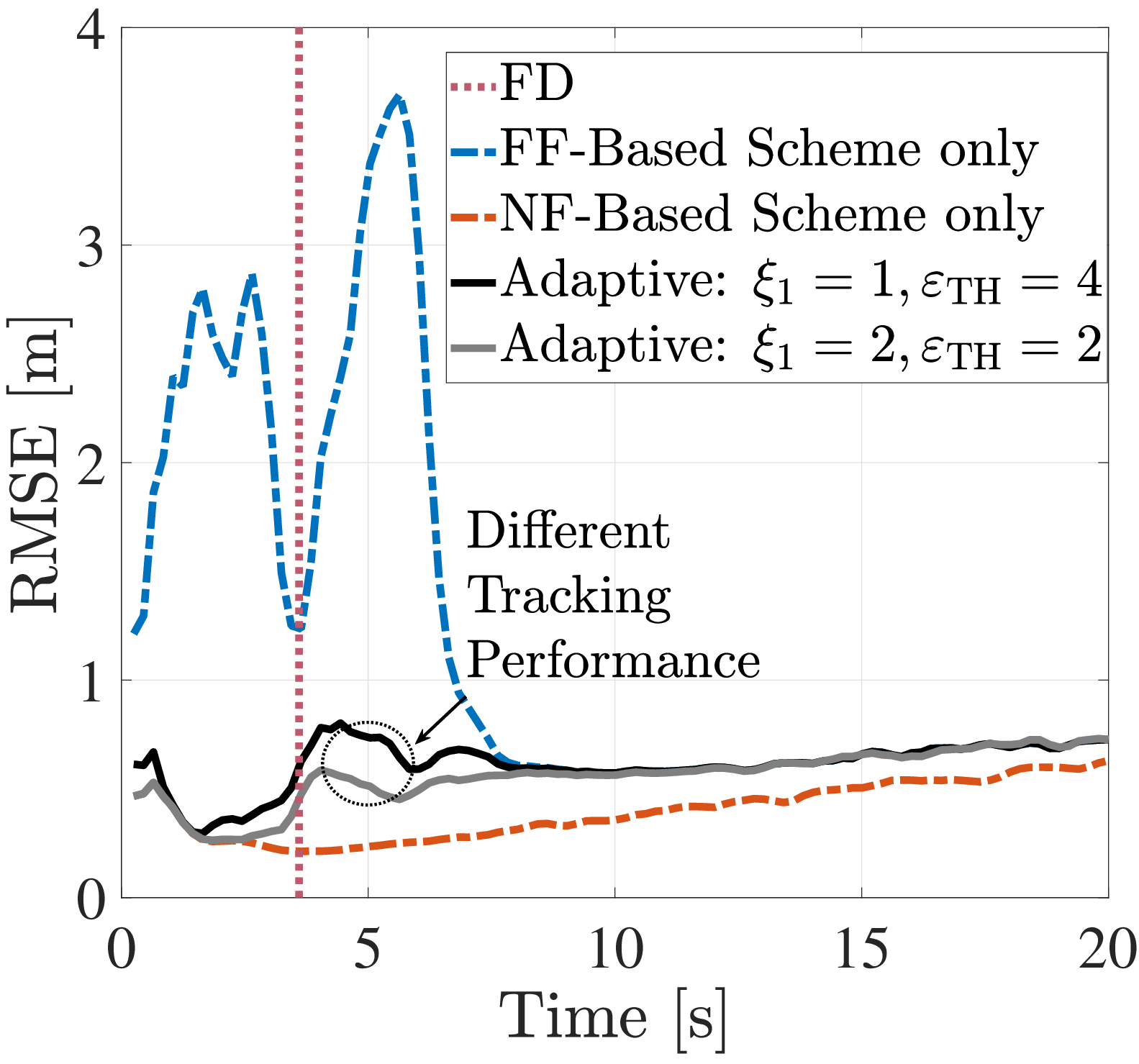}
         \caption{UE moving away from the BS at 2 m/s.}
         \label{sfig:track_rmse_b}
     \end{subfigure}
        \caption{Tracking RMSE for different implementations of the tracking procedure. The UE moves with a fixed azimuth, relative to the BS, and Scenario A is considered.}
        \label{fig:track_rmse}
\end{figure}

Figure~\ref{fig:track_rmse} shows the RMSE of the considered tracking scenario for a UE having two different trajectories. It is shown that the tracking ability of the FF-based scheme is compromised when the UE is close to the BS, while there is a performance gap between the two schemes. Using exclusively the FF-based scheme (resp. NF-based scheme) essentially marks the performance limit in lowest (resp. highest) accuracy and complexity for our proposed adaptive scheme. Our proposed adaptive procedure successfully avoids the unreliable position estimations of the FF-based scheme and approaches the performance of the NF-based scheme when necessary. Two variations of Alg.~\ref{alg:track_proc} are shown, with different initial conditions ($\xi_{1} = \{1,2\}$) and different levels of trajectory divergence tolerance ($\varepsilon_{\rm{TH}} = \{2,4\}$). In Fig.~\ref{sfig:track_rmse_a}, the UE starts in the FF and maintains the FF-based scheme for a long period as it approaches the BS, even when the initial indication is to use the NF-based scheme ($\xi_{1}=2$). When the trajectory-estimation error thresehold is low, the NF-based scheme is employed more often and the switch occurs faster. This leads to a minor performance gain, at the cost of employing the high complexity scheme in more occasions. 

In Fig.~\ref{sfig:track_rmse_b}, the UE starts in the NF and employs the NF-based scheme, even when the initial indication is to use the FF-based scheme ($\xi_{1}=1$). As the UE moves away from the BS the FF-based scheme is utilized almost exclusively. Respectively, allowing for the position estimation to diverge from the trajectory allows the low-complexity scheme to be employed more frequently, when the UE is close to the NF-FF border. Algorithm~\ref{alg:track_proc} allows for a lenient (strict) design which eventually leads to a complexity (resp. accuracy) gain. 

On average, in Fig.~\ref{sfig:track_rmse_a} when the tracking process initiated in the FF, the NF-based scheme was utilised in 20.9\% of the total position estimations for $\varepsilon_{\rm{TH}} = 4$, while, for $\varepsilon_{\rm{TH}} = 2$, it was 32.5\%. In Fig.~\ref{sfig:track_rmse_a}, when the tracking process initiated in the NF, the NF-based scheme was utilised in 21.3\% of the total position estimations for $\varepsilon_{\rm{TH}} = 4$, while, for $\varepsilon_{\rm{TH}} = 2$, it was 28.6\%.  This highlights that the performance gain of the strict design comes with the signaling overhead and computational cost of employing the NF-based scheme more often. Moreover, the initial conditions may play an important role when the UE is close to the NF-FF border. When the UE is far away, the two schemes have similar performance, while it is reliably detectable when the UE is close to the BS. In either case, the adaptive procedure quickly converges to the appropriate scheme. As discussed in Section~\ref{sec:track}, the tracking process provides various degrees of freedom in its design. Here, the accuracy of the adaptive scheme is effectively below 1 m while the low-complexity scheme is employed as much as possible. Figure~\ref{fig:track_rmse} highlights that a sense of memory utilizing prior positional information allows for successfully implementing a localization procedure of adaptive complexity.   

\begin{figure}[h]
    \centering
    \includegraphics[width=\columnwidth]{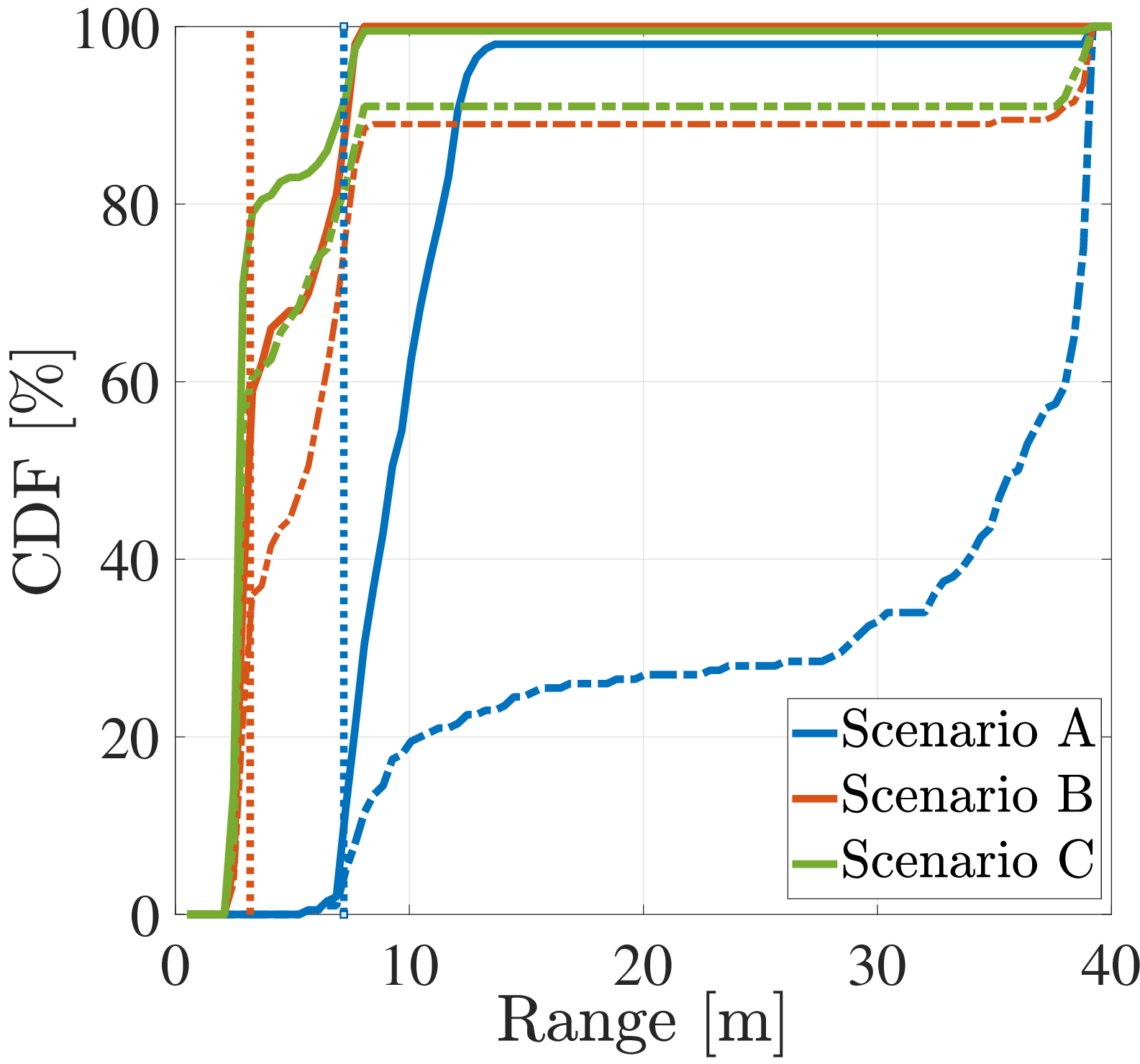}
    \caption{CDF for the switching distance for different scenarios. The UE is moving towards the BS. For the solid lines, the trajectory-estimation error thresehold is $\varepsilon_{\rm{TH}} = 4$, while for the dashed lines $\varepsilon_{\rm{TH}} = 2$. The dotted lines mark the NF region $d_{\rm{NF}} = \text{FD}$.}
    \label{fig:cdf_sd}
\end{figure}

Figure~\ref{fig:cdf_sd} shows the switching distance in a tracking application where the UE starts in the FF and moves towards the BS. Initially, the FF-based scheme is employed ($\xi_{1} = 1$) and the switching distance is defined as the first request for the NF-based scheme to be employed. It is shown that the FD distance is not necessarily a good indicator for the appropriate switching distance. Scenario A refers to a larger BS array and it it shown that the switching point has a larger variance. This implies that there is a larger \emph{uncertainty region} where the FF scheme might be unreliable. Scenarios B and C represent smaller BS arrays and the average switching distance decreases, while a large portion of UEs only switches to the NF scheme when they surpass the $d_{\rm{NF}}$ threshold. In addition, when the UE distribution is more favorable, i.e., the average $\Delta z$ is smaller in Scenario C, we observe a small percentage of UEs requesting the high complexity scheme to be employed before they enter the NF region. This highlights our intuition that the considered geometry plays a vital role in the design of a localization scheme that relies on the BS-to-UE distance. 

In~\ref{fig:cdf_sd}, the dashed lines represent a strict design, with a low leniency on diverging from the formed trajectory. As a result, the CDF shows a larger spread for all scenarios, with UEs requesting a switch to high complexity faster. A small error tolerance results in inefficient resource management, since the NF-based signaling is utilized a large area, where the FF-based scheme could provide sufficient performance. In addition, the large spread indicates that the system is susceptible to large initial trajectory error, as a few estimations are stored in memory. As we alter the illustrated design parameters, the system's behaviour changes drastically, which highlights the degrees of freedom bestowed on the system designer in Alg.~\ref{alg:track_proc}, depending on the desired accuracy, computational complexity and signaling overhead.

\section{Conclusions}
In this paper, we proposed two signaling designs and two DL localization algorithms with different signaling overheads and computational complexities. The distinctive characteristic of each scheme is the validity of the FF/NF assumption and the BS's ability to effectively cover the whole region with beam-steering rather than beam-focusing techniques. We illustrate that the considered BS-to-UE range effectively shapes the signaling requirements from the BS's side, while it dictates which algorithm needs to be employed at the UE.  We explore the trade-off between complexity and accuracy for different algorithms and we highlight that the standard definition of the FF region may be inadequate when a practical localization scheme is to be designed based on the BS-to-UE distance. Therefore, the FD distance is not necessarily a good metric for the validity of the FF/NF assumption. In addition, we have proposed an adaptive protocol for a tracking application, where previous position information validates the reliability of the low-complexity position estimation. We explored the various degrees of freedom upon designing our iterative procedure that shapes the system's behaviour in terms of accuracy and complexity. Thus, the overall complexity and signaling overhead is reduced while the tracking performance is maintained within an acceptable range. 

\bibliographystyle{ieeetr}
\bibliography{references}

\begin{thebibliography}{10}

\bibitem{ArxivAdaptiveDL}
G.~Mylonopoulos, B.~Makki, F.~G\'abor, and S.~Buzzi, ``{Adaptive Downlink Localization in Near-Field and Far-Field},'' {\em arXiv preprint arXiv: 2402.02473}, 2024.

\bibitem{de2021convergent}
C.~De~Lima, D.~Belot, R.~Berkvens, A.~Bourdoux, D.~Dardari, M.~Guillaud, M.~Isomursu, E.-S. Lohan, Y.~Miao, A.~N. Barreto, {\em et~al.}, ``Convergent communication, sensing and localization in {6G} systems: An overview of technologies, opportunities and challenges,'' {\em IEEE Access}, vol.~9, pp.~26902--26925, January 2021.

\bibitem{chen2022channel}
H.~Chen, A.~Elzanaty, R.~Ghazalian, M.~F. Keskin, R.~J{\"a}ntti, and H.~Wymeersch, ``Channel model mismatch analysis for {XL-MIMO} systems from a localization perspective,'' in {\em Proc. GLOBECOM 2022-2022 IEEE Global Communications Conference}, pp.~1588--1593, IEEE, December 2022.

\bibitem{cui2022near}
M.~Cui, Z.~Wu, Y.~Lu, X.~Wei, and L.~Dai, ``Near-field {MIMO} communications for {6G}: Fundamentals, challenges, potentials, and future directions,'' {\em IEEE Communications Magazine}, vol.~61, no.~1, pp.~40--46, September 2022.

\bibitem{jingjing2021search}
P.~Jingjing, S.~P. Raj, and M.~Shaoyang, ``A search-free near-field source localization method with exact signal model,'' {\em Journal of Systems Engineering and Electronics}, vol.~32, no.~4, pp.~756--763, August 2021.

\bibitem{su2021deep}
X.~Su, Z.~Gong, P.~Hu, T.~Liu, B.~Peng, and Z.~Liu, ``Deep unfolding network for near-field source localization via symmetric nested array,'' in {\em Proc. 2021 CIE International Conference on Radar (Radar)}, pp.~1593--1597, IEEE, December 2021.

\bibitem{shu2020near}
T.~Shu, L.~Li, and J.~He, ``Near-field source localization with two-level nested arrays,'' {\em IEEE Communications Letters}, vol.~24, no.~11, pp.~2488--2492, July 2020.

\bibitem{guanghui2019high}
C.~Guanghui, Z.~Xiaoping, J.~Shuang, Y.~Anning, and L.~Qi, ``High accuracy near-field localization algorithm at low {SNR} using fourth-order cumulant,'' {\em IEEE Communications Letters}, vol.~24, no.~3, pp.~553--557, December 2019.

\bibitem{huang2022near}
S.~Huang, B.~Wang, Y.~Zhao, and M.~Luan, ``Near-field {RSS}-based localization algorithms using reconfigurable intelligent surface,'' {\em IEEE Sensors Journal}, vol.~22, no.~4, pp.~3493--3505, January 2022.

\bibitem{ArxvivUEDetRIS}
G.~Mylonopoulos, B.~Makki, S.~Buzzi, and F.~G\'abor, ``{Joint User Detection and Localization in Near-Field Using Reconfigurable Intelligent Surfaces},'' {\em arXiv preprint arXiv: 2402.02488}, 2024.

\bibitem{yang2021communication}
J.~Yang, Y.~Zeng, S.~Jin, C.-K. Wen, and P.~Xu, ``Communication and localization with extremely large lens antenna array,'' {\em IEEE Transactions on Wireless Communications}, vol.~20, no.~5, pp.~3031--3048, January 2021.

\bibitem{pan2023ris}
Y.~Pan, C.~Pan, S.~Jin, and J.~Wang, ``{RIS}-aided near-field localization and channel estimation for the {T}erahertz system,'' {\em IEEE Journal of Selected Topics in Signal Processing}, June 2023.

\bibitem{luan2021phase}
M.~Luan, B.~Wang, Y.~Zhao, Z.~Feng, and F.~Hu, ``Phase design and near-field target localization for {RIS}-assisted regional localization system,'' {\em IEEE Transactions on Vehicular Technology}, vol.~71, no.~2, pp.~1766--1777, December 2021.

\bibitem{fascista2021downlink}
A.~Fascista, A.~Coluccia, H.~Wymeersch, and G.~Seco-Granados, ``Downlink single-snapshot localization and mapping with a single-antenna receiver,'' {\em IEEE Transactions on Wireless Communications}, vol.~20, no.~7, pp.~4672--4684, March 2021.

\bibitem{nazari2023mmwave}
M.~A. Nazari, G.~Seco-Granados, P.~Johannisson, and H.~Wymeersch, ``Mm{W}ave {6D} radio localization with a snapshot observation from a single {BS},'' {\em IEEE Transactions on Vehicular Technology}, vol.~72, no.~7, pp.~8914 -- 8928, February 2023.

\bibitem{mylonopoulos2022active}
G.~Mylonopoulos, C.~D’Andrea, and S.~Buzzi, ``Active reconfigurable intelligent surfaces for user localization in mm{W}ave {MIMO} systems,'' in {\em Proc. 2022 IEEE 23rd International Workshop on Signal Processing Advances in Wireless Communication (SPAWC)}, pp.~1--5, IEEE, July 2022.

\bibitem{mylonopoulos2023maximum}
G.~Mylonopoulos, L.~Venturino, S.~Buzzi, and C.~D’Andrea, ``Maximum-likelihood user localization in active-{RIS} empowered mm{W}ave wireless networks,'' in {\em Proc. 2023 17th European Conference on Antennas and Propagation (EuCAP)}, pp.~1--5, IEEE, March 2023.

\bibitem{de2020near}
A.~de~Jesus~Torres, L.~Sanguinetti, and E.~Bj{\"o}rnson, ``Near-and far-field communications with large intelligent surfaces,'' in {\em 2020 54th Asilomar Conference on Signals, Systems, and Computers}, pp.~564--568, IEEE, November 2020.

\bibitem{zhang20236g}
H.~Zhang, N.~Shlezinger, F.~Guidi, D.~Dardari, and Y.~C. Eldar, ``{6G} wireless communications: From far-field beam steering to near-field beam focusing,'' {\em IEEE Communications Magazine}, March 2023.

\bibitem{bjornson2021primer}
E.~Bj{\"o}rnson, {\"O}.~T. Demir, and L.~Sanguinetti, ``A primer on near-field beamforming for arrays and reconfigurable intelligent surfaces,'' in {\em Proc. 2021 55th Asilomar Conference on Signals, Systems, and Computers}, pp.~105--112, IEEE, October 2021.

\bibitem{he2021mixed}
J.~He, L.~Li, T.~Shu, and T.-K. Truong, ``Mixed near-field and far-field source localization based on exact spatial propagation geometry,'' {\em IEEE Transactions on Vehicular Technology}, vol.~70, no.~4, pp.~3540--3551, March 2021.

\bibitem{tian2021phase}
Y.~Tian, X.~Gao, W.~Liu, and H.~Chen, ``Phase compensation-based localization of mixed far-field and near-field sources,'' {\em IEEE Wireless Communications Letters}, vol.~11, no.~3, pp.~598--601, December 2021.

\bibitem{huang2020one}
Z.~Huang, W.~Wang, F.~Dong, and D.~Wang, ``A one-snapshot localization algorithm for mixed far-field and near-field sources,'' {\em IEEE Communications Letters}, vol.~24, no.~5, pp.~1010--1014, February 2020.

\bibitem{wang2017unified}
Y.~Wang and K.~Ho, ``Unified near-field and far-field localization for {AOA} and hybrid {AOA-TDOA} positionings,'' {\em IEEE Transactions on Wireless Communications}, vol.~17, no.~2, pp.~1242--1254, December 2017.

\bibitem{he2022mixed}
J.~He, T.~Shu, L.~Li, and T.-K. Truong, ``Mixed near-field and far-field localization and array calibration with partly calibrated arrays,'' {\em IEEE Transactions on Signal Processing}, vol.~70, pp.~2105--2118, April 2022.

\bibitem{yan2023improved}
H.~Yan, Y.~Wang, Y.~Gong, Z.~Zhang, and L.~Wang, ``Improved sparse symmetric arrays design for mixed near-field and far-field source localization,'' {\em IEEE Transactions on Aerospace and Electronic Systems}, July 2023.

\bibitem{guerra2021near}
A.~Guerra, F.~Guidi, D.~Dardari, and P.~M. Djuri{\'c}, ``Near-field tracking with large antenna arrays: Fundamental limits and practical algorithms,'' {\em IEEE Transactions on Signal Processing}, vol.~69, pp.~5723--5738, August 2021.

\bibitem{palmucci2023two}
S.~Palmucci, A.~Guerra, A.~Abrardo, and D.~Dardari, ``Two-timescale joint precoding design and {RIS} optimization for user tracking in near-field {MIMO} systems,'' {\em IEEE Transactions on Signal Processing}, August 2023.

\bibitem{li2020self}
Z.~Li, J.~Cao, X.~Liu, J.~Zhang, H.~Hu, and D.~Yao, ``A self-adaptive bluetooth indoor localization system using {LSTM}-based distance estimator,'' in {\em Proc. 2020 29th International Conference on Computer Communications and Networks (ICCCN)}, pp.~1--9, IEEE, August 2020.

\end{thebibliography}

\end{document}